\documentclass[reprint,physrev,nofootinbib]{revtex4-2}
\usepackage{amsmath} 
\usepackage{graphicx}
\usepackage{dcolumn}
\usepackage{bm}
\usepackage{hyperref}
\hypersetup{
    colorlinks=true,
    citecolor = blue
}

\usepackage{amsmath}
\usepackage[normalem]{ulem}

\usepackage[abs]{overpic}
\newcommand{\p}{\partial}

\newcommand{\la}{\langle}
\newcommand{\ra}{\rangle}
\newcommand{\bh}{\bar{h}}
\newcommand{\bj}{\bar{j}}
\newcommand{\bvarphi}{\bar{\varphi}}

\newcommand{\Gk}{\hat G_\kappa}
\newcommand{\Fk}{\hat F_\kappa}
\newcommand{\Ck}{\hat C_\kappa}
\newcommand{\Tk}{\hat T_\kappa}

\usepackage[dvipsnames]{xcolor}

\begin{document}

\title{\textbf{Scaling regimes of the  Kuramoto-Sivashinsky equation  from the functional renormalization group}
}

\author{Liubov Gosteva$^1$, Nicol\'as Wschebor$^2$, L\'eonie Canet$^1$}\email{Contact author: leonie.canet@lpmmc.cnrs.fr}
\affiliation{$^1$ Universit\'e Grenoble Alpes, CNRS, LPMMC, 38000 Grenoble, France   \\$^2$ Instituto de F\'isica, Facultad de Ingenier\'ia, Universidad de la Rep\'ublica, J.H.y Reissig 565, 11000 Montevideo, Uruguay}

\begin{abstract}
We revisit the renormalization group (RG) approach to the one-dimensional stochastic Kuramoto-Sivashinsky (KS) equation and show that previous approaches based on perturbative Wilsonian RG with a sharp cutoff are not valid, even though  they yield a qualitatively correct picture. The reason is that taking momentum derivatives while using the sharp cutoff   is not well-defined in some cases and leads to intrinsic divergencies. This is a well-known problem of Wilsonian RG, which can be simply cured by using a smooth cutoff, and employing  the functional renormalization group (FRG) framework.  We establish the flow equations for the KS model within the FRG, and demonstrate that it flows to the Kardar-Parisi-Zhang (KPZ) fixed point at large scales. We then calculate the full two-point correlation function over a wide range  of momenta and frequencies. We show that it exhibits three universal scaling regimes that we characterize: the KPZ regime (with dynamical exponent $z=3/2$), the Edwards-Wilkinson regime (with $z=2$) and the recently discovered  inviscid regime (with $z=1$). The latter develops over an extended range of large wavenumbers and originates from the vanishing of the effective viscosity. Lastly, we investigate the large-scale behavior of the deterministic KS equation by studying the limit of vanishing noise and we determine the scales where the KPZ regime can emerge in the deterministic case.
\end{abstract}

\maketitle

\section{Introduction}

The Kuramoto-Sivashinsky (KS) equation is the simplest partial differential equation with quadratic or cubic non-linearity exhibiting chaos~\cite{Brummitt2009}. It describes the dynamics of a real-valued scalar field $h(t,x)$ in a one-dimensional space as
\begin{equation} 
\p_t h = \nu \p_x^2 h - \tau \p_x^4 h + \frac{\lambda}{2} (\p_x h)^2\,,\qquad \nu<0
\label{eq:KS}
\end{equation}
where $\nu$ is a negative viscosity (or surface tension) and $\tau$
is positive to ensure global stability.
It was originally proposed independently by Kuramoto and co-workers to  model chemical turbulence in reaction-diffusion systems \cite{KuramotoTsuzuki1975, KuramotoTsuzuki1976, Kuramoto1978}, and by Sivashinsky and co-workers to describe hydrodynamic instabilities  in laminar flame fronts~\cite{Sivashinsky1977_1, MichelsonSivashinsky1977_2, Sivashinsky1977, Sivashinsky1980}. The KS equation then turned out to emerge in a great variety of physical phenomena exhibiting chaos and instabilities.
 A prominent example is  the complex Ginzburg-Landau equation (CGLE), where the KS equation appears as the effective dynamics of the phase of the complex field in the phase turbulence regime~\cite{Kuramoto1978,Shraiman1992,Aranson2002}. This weaves a deep connection with the realm of driven-dissipative Bose-Einstein condensates~\cite{Wouters2015,Vercesi2024}.

The essential feature of the deterministic KS equation lies in its negative viscosity, which generates an intrinsic modulational instability of the linearized equation, and induces a chaotic behavior when the system size is large compared to the typical scale of the instability pattern~\cite{Hyman1986,Papageorgiou1991}.
This chaotic dynamics leads to a steady-state, whose large-scale properties are believed to belong to the celebrated Kardar-Parisi-Zhang (KPZ) universality class, which means that it is effectively described at large scales by the KPZ equation \cite{Kardar1986}
\begin{equation}
\p_t h = \nu \p_x^2 h + \frac{\lambda}{2} (\p_x h)^2 + \eta\,, \qquad \nu>0
\label{eq:KPZ}
\end{equation}
with a positive viscosity $\nu$ and a white Gaussian noise $\eta(t,x)$ with zero mean and covariance
\begin{equation}
\la\eta(t,x)\eta(t',x')\ra = 2 D\,\delta(t-t')\delta(x-x')\,.
\label{eq:etaKPZ}
\end{equation}
The intrinsic instabilities of the KS equation thus generate an effective stochastic noise  and a positive effective viscosity. Since the KPZ dynamics is emergent,  we denote $\nu_{\rm eff}$, $D_{\rm eff}$ and $\lambda_{\rm eff}$ the corresponding effective values of the KPZ parameters, which are not the microscopic ones. 
 The emergence of KPZ universality was conjectured long ago  \cite{Fujisaka1977,Zaleski1989}, in particular on the basis of  renormalization group (RG) approaches \cite{Yakhot1980, Ueno2005}.
However, in early numerical studies \cite{Hyman1986,Zaleski1989, Sneppen1992, Hayot1993} another scaling regime -- the Edwards-Wilkinson (EW) regime -- was observed, due to the finite size of the system. 
The latter is characterized by a critical dynamical exponent $z=2$ corresponding to diffusive behavior, different from the KPZ value $z=3/2$
 in one dimension. The onset of the KPZ regime was roughly estimated in Ref. \cite{Sneppen1992, Hayot1993} for typical length- and time-scale given by
\begin{equation}
L\sim 152 \nu_{\rm eff}^3 / (D_{\rm eff} \lambda_{\rm eff}^2) \,, \quad
t\sim \nu_{\rm eff}^5 / (D_{\rm eff}^2 \lambda_{\rm eff}^4)\,,
\label{eq:estimationsSneppen}
\end{equation}
which are larger than the ones which were used in simulations at that time. 
 Clear numerical evidence of the KPZ scaling in the deterministic KS equation was obtained only recently \cite{Roy2020} in simulations involving very large system size and time.
More recently, another scaling regime, called inviscid scaling and characterized by $z=1$, was reported in Ref.~\cite{KSI}, discovered within the functional renormalization group (FRG)  and large-scale direct numerical simulations (DNS). Evidence of this regime was also obtained in the simulations of the CGLE \cite{Vercesi2024}, in a regime where the CGLE can be reduced to the KS description.

In this paper, we revisit the RG approaches~\cite{Yakhot1980,Cuerno1995RG,Ueno2005} to the KS equation, which are based on the stochastic KS equation, {\it i.e.} the KS equation  in the presence of a Gaussian noise
\begin{align} \label{eq:SKS}
&\p_t h = \nu \p_x^2 h - \tau \p_x^4 h + \frac{\lambda}{2} (\p_x h)^2 + \xi,  \\
&\la\xi(t,x)\xi(t',x')\ra = 2D_{\rm SKS}\delta(t-t')\delta(x-x')\,.\label{eq:SKSnoise}
\end{align}
We show that within the Wilsonian RG with a sharp cutoff, the flow equations of the parameter  $\tau$ is not well-defined. This is because its derivation, in Fourier space, involves the fourth derivatives with respect to $p$. However, in the usual Wilsonian RG, a sharp cutoff is used to separate the low and high momentum modes, and taking momentum derivatives leads to divergencies. This is a well-known drawback of the standard Wilsonian formulation which employs a sharp cutoff (see, for example, \cite{Morris:1995af}).
 The solution to this issue is to replace the non-analytical sharp cutoff with a smooth function (see,  for example, \cite{Polchinski:1983gv}). This is at the heart of the FRG formalism \cite{Wetterich1993,Ellwanger:1993kk,Morris:1993qb,Berges2002} (for a recent review, see \cite{Dupuis2021}; for a pedagogical introduction, see \cite{Delamotte2012}). It was shown in, for example Refs.~\cite{Morris:1994ie,Berges2002,Canet:2002gs,Canet:2003qd,Balog:2019rrg,DePolsi:2020pjk}), that the FRG indeed  allows one to obtain well-defined flow equations even when momentum derivatives are involved.
We  first establish the FRG flow equations for the KS dynamics within the simplest approximation,  usually denoted Derivative Expansion (DE). This allows us to take the limit of sharp cutoff and show how the divergencies emerge in this limit. With a smooth cutoff, the flow equations are well-defined and can be integrated numerically. We show that the flow reaches the KPZ fixed point in the IR limit.

\begin{figure}
\begin{overpic}[width=0.4\textwidth]{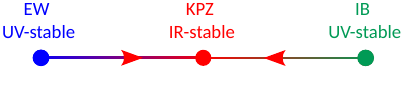}
\put(9,5){\textcolor{blue}{$\lambda=0$}}
\put(9,-2){\textcolor{blue}{$g_{\rm KPZ}=0$}}
\put(170,5){\textcolor{ForestGreen}{$\nu=0$}}
\put(170,-2){\textcolor{ForestGreen}{$g_{\rm KPZ}=\infty$}}
\end{overpic}
\caption{Fixed points of the 1D KPZ equation, with $g_{\rm KPZ}=\lambda^2 D/\nu^3$.}
\label{fig:fp}
\end{figure}
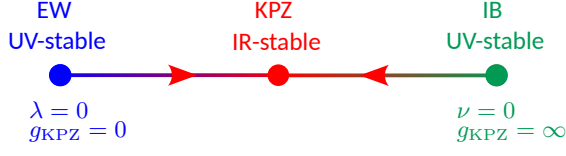

We then turn to a richer approximation, called NLO (next-to-leading order) approximation \cite{Kloss2012}, which enables us to calculate the full two-point correlation function over a broad range of frequencies and momenta. We unveil the rich universal behavior of the KS equation, by showing the existence of  three scaling regimes: KPZ, EW and IB (inviscid Burgers). These three regimes are inherited from the RG fixed-point structure of  the 1D KPZ equation,
which is depicted in Fig.~\ref{fig:fp}.
It features three fixed points:
\begin{itemize}
  \item The KPZ fixed point characterized by $z=3/2$: it is  fully attractive (stable), and  controls  the large scales (infrared (IR) behavior);
  \item The EW fixed point ($\lambda=0$, $g_{\rm KPZ}=0$)  characterized by $z=2$: it is repulsive (unstable), and controls the small scales (ultraviolet (UV) behavior) when $\lambda$ is small but non-zero, while controlling all scales when $\lambda=0$;
  \item The IB fixed point ($\nu=0$, $g_{\rm KPZ}\to\infty$)  characterized by $z=1$:
  it is repulsive (unstable), and controls the small scales when $\nu$ is small but non-zero, while controlling all scales  when $\nu=0$,
\end{itemize}
where  $g_{\rm KPZ}=\lambda^2 D/\nu^3$ is the coupling constant of the KPZ equation related to the nonlinearity.

The EW scaling is observed for momenta larger than the momentum scale set by  $g_{\rm KPZ}$, which has dimension of inverse length. This implies that if the system size is small compared to $1/g_{\rm KPZ}$,
 EW scaling is observed instead of the KPZ one as a finite-size effect~\cite{Nattermann92}. At smaller $\lambda$, the EW scaling persists over larger scales. This suggests that, for the deterministic KS equation ($D_{\rm SKS} \to 0$), since the noise is only dynamically generated, very large system sizes should be required for the KPZ scaling to set in.

The IB scaling $z=1$ was reported in simulations for the KPZ equation (or the equivalent Burgers equation) in the limit of vanishing viscosity \cite{Majda2000, Cartes2022, RodriguezFernandez2022}. It was then shown to correspond to a new fixed point of the KPZ equation, the IB fixed point \cite{Fontaine2023Unpredicted, Gosteva2024Burgers, Gosteva2025KPZtwogrids}.
The appearance of the IB regime is very natural in the KS equation and does not even require fine-tuning of parameters (such as choosing small enough  $\nu$ in the KPZ equation): since the microscopic (bare) viscosity is negative, and the effective macroscopic (renormalized) one is positive, an intermediate scale should exist where the effective viscosity vanishes. One thus expects  the IB scaling to dominate the intermediate range of wavenumbers.
Note that the existence of an unstable fixed point for the KS equation corresponding to  $\nu=0$ was pointed out in a perturbative RG study \cite{Cuerno1995RG}, although the results of this study are under question \cite{Ueno2005}. The emergence of the IB regime was clearly demonstrated in Ref.~\cite{KSI}, both through FRG calculation and extensive numerical simulations. We here explain in details the FRG calculation within the NLO
 approximation, and show that it enables one to precisely identify and locate the momentum regions of the three scaling regimes, KPZ, EW, and IB, produced by the KS dynamics.

Lastly, we address the question of the deterministic limit of the stochastic KS equation. When introducing a stochastic noise, the KPZ regime sets in for smaller system size (or equivalently at larger wavenumber), as was shown in~\cite{Ueno2005,Vercesi2024}.
This raises the question of the fate of the KPZ regime when the noise amplitude  $D_{\rm SKS}$ tends to zero. Our results indicate that the scale at which the KPZ regime sets in tends to infinity as $D_{\rm SKS}\to 0$. This indicates that, in the thermodynamic limit, the true IR regime  is the EW one.  This is not in contradiction with direct numerical simulations where the KPZ regime was observed, since  the finite  computational precision inevitably introduces some small numerical noise.

The remainder of the paper is organized as follows. We set the basis of the FRG formalism in Sec.~\ref{sec:FRG}. We then derive in Sec.~\ref{sec:LPA} the flow equations for the KS model in the simplest DE approximation, which allows us to compare with the results of perturbative RG. We finally establish that the KS flow reaches the KPZ fixed point. We implement in Sec.~\ref{sec:NLO}
 the NLO approximation to calculate the full two-point correlation function, and characterize the three scaling regimes. Sec.~\ref{sec:noise} is devoted to the study of the zero-noise $D_{\rm SKS}\to0$ limit of the stochastic KS equation to infer the behavior of the deterministic KS equation. Technical details are expounded in the Appendices.

\section{The FRG formalism}
\label{sec:FRG}

The FRG formalism is constructed from the path-integral formulation of the stochastic dynamics. This formulation can be straighforwardly  derived following the Martin-Siggia-Rose-Janssen-de Dominicis procedure~\cite{Martin1973,Janssen1976,Dominicis1976}. One obtains for the stochastic KS equation \eqref{eq:SKS}
\begin{align}
{\cal Z}[j,\bj] &= \int{\cal D}h\int{\cal D}\bh\,e^{-{\cal S}_{\Lambda}[h,\bh] +\int_{t,x}\{j h+\bj\bh\} }\nonumber\\
{\cal S}_{\Lambda}[h,\bh] &= \int_{t,x} \Big\{ \bh(t,x) {\cal E}[h(t,x)] - D_{\text{SKS}}\bh(t,x)^2 \Big\} \,\label{eq:Sbare}
\end{align}
where  ${\cal E}[h]= \p_t h - \nu \p_x^2 h + \tau \p_x^4 h - (\lambda/2) (\p_x h)^2$ is the deterministic part of the equation of motion, $\bh(t,x)$ is the response field, and  $j,\bj$ are the sources linearly coupled to the fields. This implies that ${\cal Z}[j, \bj]$ (resp. ${\cal W}=\ln {\cal Z}$) plays the role of the generating functional of   correlation functions (resp. connected correlation functions), which means that the corresponding correlations can be obtained by taking functional derivatives of  ${\cal Z}$ (resp. ${\cal W}$) with respect to the sources $j$ and $\bj$ and then setting them to zero. We use the shorthand notation $\int_{t,x}\equiv \int_{-\infty}^\infty dt\int_{-\infty}^\infty dx$.

The FRG method consists in introducing the smooth regulator term
\begin{align}
\Delta \mathcal{S}_\kappa = -\int_{t,x,x'}&\Big\{\bh(t,x) R^\nu_\kappa(x-x') \nabla'^2 h(t,x')\nonumber\\
 &+ \bh(t,x) R^D_\kappa(x-x') \bh(t,x') \Big\}
\end{align}
into the functional integral, defining the scale-dependent action
\begin{equation}
{\cal S}_\kappa = {\cal S}_\Lambda + \Delta \mathcal{S}_\kappa \,,
\label{eq:Skappa}
\end{equation}
where $\kappa$ is the RG momentum scale. It varies from $\kappa=\Lambda$, which denotes the microscopic (UV) scale at which the continuous theory is defined (or the inverse lattice spacing for a discrete model) to $\kappa=0$
 which corresponds to the macroscopic (IR) scale (infinite system length).

In the presence of the regulator, the generating functionals ${\cal Z}$ and ${\cal W}$ also become scale dependent
$\mathcal{Z}_\kappa[j,\bj] = \int \mathcal{D} h \mathcal{D} \bh\, \exp\left(-{\cal S}_\kappa[h,\bh] + \int_{t,x}\{j h + \bj \bh\} \right)$, and ${\cal W}_\kappa = \ln \mathcal{Z}_\kappa$. One then defines the effective average action as
\begin{equation}
\Gamma_{\kappa}[\varphi , \bvarphi]= -\mathcal{W}_\kappa + \int_{t,x} \big\{j \varphi + \bj \bvarphi\big\} - \Delta \mathcal{S}_\kappa\,,
\end{equation}
where $\varphi$ and $\bvarphi$ are the average fields
\begin{equation}
 \varphi(t,x) = \frac{\delta {\cal W}_\kappa}{\delta j(t,x)}\,, \qquad \bvarphi(t,x) = \frac{\delta {\cal W}_\kappa}{\delta \bj(t,x)}\,.
\end{equation}
The evolution of $\Gamma_{\kappa}$ with the RG scale is given by the exact
 Wetterich equation \cite{Wetterich1993,Ellwanger:1993kk,Morris:1993qb}
\begin{equation}  
    \p_{\kappa} \Gamma_{\kappa} =
    \frac{1}{2}\textrm{tr}\,\int_{\omega,p}
    \p_{\kappa} \mathcal{R}_{\kappa}\,
    \mathcal{G}_{\kappa}\,,
\label{eq:Wetterich}
\end{equation}
where the  trace means summation over all fields $\varphi$, $\bvarphi$,
\begin{equation}  
\mathcal{G}_{\kappa} \equiv \left(
    \Gamma_\kappa^{(2)} + \mathcal{R}_\kappa
\right)^{-1}
\label{eq:WetterichPropag}
\end{equation}
is the propagator matrix, with $\Gamma_{\kappa}^{(2)}$  the Hessian of  $\Gamma_\kappa$, and $\mathcal{R}_\kappa$ is the regulator matrix
which reads in Fourier space
\begin{equation}  
\mathcal{R}_\kappa \equiv \begin{pmatrix}
0 & p^2R^\nu_\kappa(p) \\
p^2R^\nu_\kappa(p) & -2R^D_\kappa(p) 
\end{pmatrix}\,.
\label{eq:regMatrix}
\end{equation}
More generally, one can define 1-PI (one-particle irreducible) correlation functions, also called vertices, by taking functional derivatives of $\Gamma_{\kappa}$ with respect to the average fields
\begin{align}
 \Gamma^{(m,n)}_{\kappa}&(t_1,x_1,\cdots,t_{m+n},x_{m+n}) =\nonumber\\
 &\frac{\delta^{m+n}\Gamma_{\kappa}}{\delta \varphi(t_1,x_1),\cdots, \delta \bvarphi(t_{m+n},x_{m+n})}\Big|_{\varphi=0,\bvarphi=0}\,,
\end{align}
where by convention the first $m$ derivatives are with respect to $\varphi$ and the $n$ last with respect to $\bvarphi$.

The precise form of the regulators $R^{\nu,D}_\kappa(p)$ is unimportant provided they satisfy  the following properties in Fourier space:
\begin{itemize}
  \item they are large $R^{\nu,D}_\kappa(p)\sim \kappa^2$ at small momenta $p\lesssim\kappa$, such that the slow modes are frozen and do not contribute to the functional integration in ${\cal Z_\kappa}$;
  \item they vanish $R^{\nu,D}_\kappa(p)\simeq 0$ at large $p\gtrsim\kappa$,  such that the fast modes are unaffected and are integrated in;
  \item they vanish  $R^{\nu,D}_\kappa(p)\to 0$ when  $\kappa\to0$ such that the full theory, including all fluctuation modes is recovered in this limit $\Gamma_0\equiv \Gamma$;
  \item they are very large $R^{\nu,D}_\kappa(p)\sim\Lambda^2$ when $\kappa\to\Lambda$ such that at  the microscopic scale, all fluctuation modes are frozen, and the effective average action coincides with the bare action $\Gamma_\Lambda\equiv {\cal S}_\Lambda$.
\end{itemize}
The regulator hence achieves a smooth separation of the fluctuation modes, allowing for the progressive averaging over the fast  modes, to build the effective theory for the slow ones. Owing to this construction, the effective average action $\Gamma_\kappa$  smoothly interpolates between the bare action (\ref{eq:Sbare}) at $\kappa=\Lambda$  and the effective large-lengthscale description at $\kappa=0$ that contains all the statistical properties of the system. The FRG framework thus realizes the Wilsonian RG program, replacing the sharp separation of modes with a smooth one. The effective theory  is provided by the effective average action $\Gamma_\kappa$ at each momentum scale $\kappa$.

The FRG flow equation (\ref{eq:Wetterich}) is exact, however the flow equations of the vertices $\Gamma^{(m,n)}_\kappa$ contain higher-order ones in the rhs. A closure is thus needed, allowing one to solve  these flow equations within some approximation. There exist several well-established approximation schemes within the FRG formalism~\cite{Berges2002,Delamotte2012,Dupuis2021}. In this paper, we  resort to simple ones that can be formulated as  ansatzes for the effective average action. In particular, we use: (1) a simple DE ansatz  in Sec.~\ref{sec:LPA}, to compare with the results of \cite{Ueno2005}, (2) the NLO approximation tailored to the specifics of the KPZ and KS equations in Sec.~\ref{sec:NLO}, and (3) a modification of the DE, denoted $p_0$-DE, conceived to appropriately handle a singularity  appearing at low noise values in Sec.~\ref{sec:noise}. We consider a homogeneous isotropic system throughout the paper, and we describe the system in the stationary regime, so that the  influence of initial conditions can be safely  neglected.

\section{DE approximation and comparison with perturbative RG}
\label{sec:LPA}

\subsection{Simplest DE ansatz}

The simplest ansatz for the effective action  consists in promoting the parameters $\nu$, $\tau$,  and $D_{\rm SKS}$ to scale-dependent ones $\nu_\kappa$, $\tau_\kappa$, $D_\kappa$. This corresponds to the lowest order of the derivative expansion~\cite{Carlosfuture}, as we explain below. Prior to this, let us first discuss on general grounds the domain of validity of the DE and its precision level.

The DE is an approximation scheme designed to study the small-wavenumber regime. In this context, ``small wavenumbers'' refers to the study of the full set of vertices associated with a model when all wavenumbers are less than or of the order of the inverse of the  correlation length of the problem \cite{Blaizot2006,Benitez2009}. In a critical system, such as the KPZ or KS equations, this corresponds to vertices (or their derivatives) for zero wavenumbers. This allows for the determination of the phase diagram of a model or  the calculation of most of its critical exponents, but it is important to keep in mind that the DE does not allow for the analysis of the vertices of a critical theory at nonzero wavenumbers. In the present case,  the DE should in principle be appropriate, if one is interested in the equivalence between two critical models, since this aspect can be analyzed in the small-wavenumber regime.

 It is usually not possible to establish an approximation scheme exclusively for the small-wavenumber regime, since most field theory techniques (such as the Schwinger–Dyson equations) relate vertices with certain momenta to vertices with arbitrarily different momenta. In this regard, the Wilson RG equations or the FRG equations are better suited  than the former techniques, since the wavenumber regime $p_i\simeq \kappa$ of any vertex is coupled only to other vertices that also have wavenumbers satisfying $p_i\simeq \kappa$ \cite{Morris:1994ie,Morris:1999ba,Morris:2000hm,Blaizot2006}. This forms the basis for establishing a controlled expansion in  momenta in such contexts. Furthermore, in the more specific framework of the FRG, due to the 1-PI nature of the vertices present in the Wetterich equation \cite{Wetterich1993,Ellwanger:1993kk,Morris:1993qb}, this wavenumber expansion is associated, for a wide variety of models, with a small parameter $\sim 1/4$ that controls the relative size of successive orders. This property is very general and has been robustly established and tested in various problems of equilibrium statistical mechanics \cite{Balog:2019rrg,DePolsi:2020pjk,DePolsi:2021cmi,DePolsi:2022wyb,Chlebicki:2022pxm,Sanchez-Villalobos:2024vmd,DePolsi:2025gqj}. Its validity out of equilibrium has not been established with the same level of rigor, but it has already proven to be very efficient in other non-equilibrium settings (in the domain of small wavenumbers)~\cite{Canet2004, Canet2005, Benitez13, Gredat14}.

Let us now discuss how one can apply the DE to the KPZ/KS problem. First, note that in general, in  the DE, one considers  full field-dependent renormalization functions \cite{Morris:1994ie,Berges2002,Canet:2002gs,Canet:2003qd,Balog:2019rrg,DePolsi:2020pjk}, which is not what is implemented here. 
However, when taking into account three very important symmetries that are present in the one-dimensional KPZ equation: the time-shift symmetry, the time-dependent Galilean invariance and the time-reversal symmetry, one concludes that the $\bvarphi$ and time derivatives both must be counted as two space derivatives. Moreover, due to the time-shift symmetry, the field $\varphi$ can never appear without derivatives. As a consequence, in the KPZ universality class there is no field-dependent running functions in the DE but only running coupling constants (for a more detailed analysis of this point, see \cite{Carlosfuture}). At the leading order of this counting, the terms appearing in the KPZ action (which all counts as four derivatives) are the only ones that can be included at order $\partial^4$ of the DE. This does not include the extra KS term proportional to $\tau$, which is of order $\partial^6$. Accordingly, one must at least extend the leading DE ansatz to incorporate this term. Moreover, as was noticed in \cite{Ueno2005}, in order to enable the flow to reach the KPZ regime, one should  allow for one more flowing parameter, denoted $\zeta_{\kappa}$, which is generated by the KS flow. This parameter is introduced in the noise correlator, replacing  $D_\text{SKS}$ in (\ref{eq:SKSnoise}) with $D_\kappa-\zeta_{\kappa}\p_x^2$, its role is explained in the following. With all these considerations, the ansatz in the simplest DE approximation hence writes
\begin{align}
\Gamma_{\kappa}[\varphi,\bvarphi]&= \int_{t,x} \bvarphi \Big\{
    \p_t \varphi - (\nu_\kappa-\tau_\kappa\p_x^2)\p_x^2\varphi-\frac{\lambda}{2}(\p_x\varphi)^2
    \nonumber \\
    &-(D_\kappa-\zeta_{\kappa}\p_x^2)\bvarphi
\Big\} \,.
\label{eq:Gammak}
\end{align}
This form is compatible with the extended symmetries of the KS action and satisfy all the corresponding Ward identities \cite{Kloss2012}. In details, the KS equation shares with the KPZ equation two fundamental symmetries, which are  the time-dependent Galilean and shift symmetries (more details can be found in Refs.~\cite{Canet2011,Kloss2012}). They impose in particular that the parameter $\lambda$ and the prefactor 1 in front of $\p_t \varphi$ are not renormalized and remain equal to their bare value.  With this ansatz, the two-point vertices in Fourier space are given by
\begin{align}
\Gamma^{(1,1)}_{\kappa}(\omega,p) &= i \omega + p^2(\nu_\kappa+\tau_\kappa p^2)\,,
\nonumber \\
\Gamma^{(0,2)}_{\kappa}(\omega,p) &=-2(D_\kappa+ \zeta_{\kappa}p^2)\,.
\label{eq:Gamma2}
\end{align}

Let us now comment on the $\zeta_\kappa$ parameter. 
As stated before, the 1D KPZ equation satisfies an additional symmetry which is a time-reversal symmetry, and which yields the exact fluctuation-dissipation relation \cite{Frey94,Canet2011}
\begin{equation}
2\,\text{Re}\Gamma^{(1,1)}_\kappa(\omega,p) = -\frac{\nu}{D}p^2\Gamma^{(0,2)}_\kappa(\omega,p)\,,
\label{eq:FDT}
\end{equation}
where $\nu$ and $D$ are the bare parameters entering the KPZ equation \eqref{eq:KPZ}.
In the KS equation, the time-reversal symmetry is not present at the microscopic level, but it is expected to emerge at a scale in the IR, denoted  by $\kappa_{\rm TRS}$, through the renormalization process if the flow leads to the KPZ fixed point\footnote{Strictly speaking,  the time-reversal symmetry is realized only asymptotically, but fixing a certain level of precision, one can determine a value of $\kappa_{\rm TRS}$ below which time  reversal symmetry is satisfied.}. In this case, the bare values $\nu$, $D$ in relation \eqref{eq:FDT} should be replaced by their renormalized values at this scale $\nu_{\kappa_{\rm TRS}}$, $D_{\kappa_{\rm TRS}}$.
Therefore, within the simplest ansatz~\eqref{eq:Gammak}, the relation (\ref{eq:FDT}) imposes
\begin{align}
\nu_\kappa = \alpha D_\kappa\,,
\quad
\tau_\kappa = \alpha \zeta_{\kappa}\,,
\quad
\alpha \equiv \frac{\nu_{\kappa_{\rm TRS}}}{D_{\kappa_{\rm TRS}}}\,,
\label{eq:FDT1}
\end{align}
and this should hold  along the flow for $\kappa \leq \kappa_{\rm TRS}$ up to the KPZ fixed point. Without introducing $\zeta_{\kappa}$, the KPZ fixed point would still be reached, but with an only approximate time-reversal symmetry, yielding that the dynamical exponent would be close to, but not exactly equal to $z=3/2$. We therefore include the flow of $\zeta_{\kappa}$. The initial condition for the flow equations at $\kappa=\Lambda$ is $\nu_\Lambda=\nu$, $\tau_\Lambda=\tau$, $D_\Lambda=D_{\rm SKS}$, $\zeta_{\Lambda}=0$. Even if initially zero, the latter parameter is generated by the flow.

The propagator matrix entering the Wetterich equation is given by
\begin{equation}
{\cal G}_\kappa(\omega,p)= \begin{pmatrix}
C_\kappa(\omega,p) & G_\kappa(-\omega,p)   \\
 G_\kappa(\omega,p)  & 0
\end{pmatrix}\, ,
\label{eq:propag}
\end{equation}
where $C_\kappa$ is the connected correlation function and $G_\kappa$ is the connected  response function:
\begin{align}
C_\kappa(\omega,p) &\equiv 
\mathcal{F} [\langle h(t,x) h(0,0)\rangle_{\text{c}}]\,,\\
G_\kappa(\omega,p) &\equiv 
\mathcal{F} [\langle \bh(t,x) h(0,0)\rangle_{\text{c}}]\,.
\label{eq:corr}
\end{align}
Within the simple DE approximation, one has
\begin{align}
C_\kappa(\omega,p) &= \frac{2(D_\kappa+\zeta_{\kappa}p^2+R_\kappa^D(p))}{\omega^2 + p^4(\nu_\kappa+\tau_\kappa p^2+R_\kappa^\nu(p))^2}\,,\nonumber\\
G_\kappa(\omega,p)&= \frac{1}{i \omega + p^2(\nu_\kappa+\tau_\kappa p^2+R_\kappa^\nu(p))}\,,
\label{eq:propDE}
\end{align}
and the only higher-order vertex within this approximation is the bare three-point one given by
\begin{equation}
 \Gamma_\kappa^{(2,1)}(\omega_1,p_1,\omega_2,p_2) = \lambda p_1 p_2\,.
 \label{eq:gamma21}
\end{equation}

\subsection{Dimensionless variables and flow equations}

Within the RG formalism, scale invariance ({\it i.e.} independence of any scale) corresponds to a fixed point ({\it i.e.} to a stationary point of the RG equation $\p_\kappa(\cdot) =0$). To study scaling regimes, it is thus convenient  to introduce dimensionless variables.
One defines dimensionless momenta as $\hat p = p/\kappa$. For the KPZ flow, one usually defines dimensionless frequencies as  $\hat \omega = \omega/(\nu_\kappa\kappa^2)$. However, this choice is not suitable for the KS flow as $\nu_\kappa$ is expected to vanish at some scale.
In Ref.~\cite{Ueno2005}, the adimensionalization is achieved using  $\tau_\kappa$ instead of $\nu_\kappa$. However, we find that  $\tau_\kappa$ also crosses zero, starting from a positive value corresponding to the microscopic value in the KS equation, and attaining a small negative value in the IR. This is in accordance with the shape of the full  momentum-dependent effective viscosity (see Sec.~\ref{sec:NLO}). Thus, we choose to adimensionalize using $D_\kappa$ and $\kappa$. We hence define  the dimensionless frequencies as $\hat \omega \equiv \omega/(\kappa^2 D_\kappa)$, and the four dimensionless parameters\footnote{
    Note that the parameters are dimensionless in the RG sense ({\it i.e} referring to {\it scaling} dimensions): in the IR, at $\kappa\rightarrow 0$, they stop to evolve and reach a fixed point plateau. For example, the dimensionless variable $\Fk$ indicates that $\nu_\kappa$ and $D_\kappa$ scale in the same way in the IR, in line with \eqref{eq:FDT1}. The measurement units should be  accounted for separately from the adimensionalization \eqref{eq:UenoLPAdimless}.
}:
\begin{equation}
\Gk=\lambda^2\frac{D_\Lambda}{\kappa D_\kappa^2}
\,,\;
\Fk=\frac{\nu_\kappa}{D_\kappa}
\,,\;
\Ck=\frac{\kappa^2 \zeta_{\kappa}}{D_\kappa}
\,,\;
\Tk=\frac{\kappa^2 \tau_\kappa}{D_\kappa}\,,\label{eq:UenoLPAdimless}
\end{equation}
which evolve according to the dimensionless flow equations
\begin{align}
\p_s \Gk &= (2\eta^D_\kappa-1) \Gk\,, \label{eq:flowG}
\\
\p_s \Fk &= \eta^D_\kappa \Fk + \frac{\p_s \nu_\kappa}{D_\kappa}\,, \label{eq:flowF}
\\
\p_s \Ck &= (\eta^D_\kappa+2) \Ck + \frac{{\kappa^2}\p_s  \zeta_{\kappa}}{ {D_\kappa}}\,, \label{eq:flowC}
\\
\p_s \Tk &= (\eta^D_\kappa+2) \Tk + \frac{\kappa^2\p_s \tau_\kappa}{D_\kappa}\,,\label{eq:flowT}
\end{align}
where 
\begin{equation}
\eta^D_\kappa \equiv -\p_s D_\kappa/D_\kappa\label{eq:etaD}
\end{equation}
is the anomalous dimension, and $s\equiv \ln (\kappa/\Lambda)$ is the ``RG time''.

We parametrize the  regulators in the following form:
\begin{equation}
R_\kappa^D(p^2) = D_\kappa r_D(\hat p^2)\,,\quad R_\kappa^\nu(p^2) = D_\kappa r_\nu(\hat p^2)\,. \label{eq:regLPA}
\end{equation}
For a symmetry to be preserved,  $\Gamma_\kappa^{(2)}+{\cal R}_\kappa$ should satisfy this symmetry. In particular,  the fluctuation-dissipation  relations (\ref{eq:FDT}), (\ref{eq:FDT1})  are verified if one imposes
\begin{equation}
r_\nu(\hat p^2)=\alpha r_D(\hat p^2) \equiv \alpha r(\hat p^2)\,, 
\label{eq:regLPAalpha}
\end{equation}
Hence, we make this choice of regulators, with constant ratio $\alpha$, which is compatible with an emergent time-reversal symmetry.

The flow equations (\ref{eq:flowG})-(\ref{eq:flowT}) consist of a part linear in the dimensionless parameter and a nonlinear part. The latter can be obtained from the dimensionful flows of the two-point vertices as follows
\begin{align}
 \p_s D_\kappa &= -\frac 1 2 \p_s\Gamma^{(0,2)}_{\kappa}(0,0) \nonumber\\
 \p_s \nu_\kappa &= \frac 1 2  \left[\frac{\p}{\p p^2}\p_s\Gamma^{(1,1)}_{\kappa}(0,p)\right]\Big|_{p=0} \nonumber\\
 \p_s \zeta_{\kappa} &= -\frac {1}{4}  \left[\frac{\p}{\p p^2}\p_s\Gamma^{(0,2)}_{\kappa}(0,p)\right]\Big|_{p=0} \nonumber\\
  \p_s \tau_\kappa &= \frac{1}{24} \left[\frac{\p}{\p p^4} \p_s\Gamma^{(1,1)}_{\kappa}(0,p)\right]\Big|_{p=0} \label{eq:defparam}
\end{align}
where the flows of $\Gamma^{(1,1)}$ and  $\Gamma^{(0,2)}$ are computed from the Wetterich equation~\eqref{eq:Wetterich}, using the propagator~\eqref{eq:propag}, \eqref{eq:propDE} and vertex~\eqref{eq:gamma21} corresponding to the DE ansatz.

This gives, after integration over the internal (loop) frequency $\omega$, the following flow equations (which are available in \cite{supmat_nb} in a Mathematica~\cite{Mathematica} notebook):
\begin{align}
&\frac{\p_s D_\kappa}{D_\kappa} = 
-\Gk\int_{\hat q} \frac{h_D (3 s_\nu h_D - 2 s_D h_\nu)}{4 {\hat q}^2 h_\nu^4},\label{eq:dsD}
\\
&\frac{\p_s \nu_\kappa}{D_\kappa} =
-\Gk\int_{\hat q} \frac{ 
    2 {\hat q}^2 \left( 
        s_D h_\nu^\prime-s_\nu h_D^\prime
    \right)
    + h_D s_\nu
}{4 {\hat q}^2 h_\nu^3},\label{eq:dsnu}
\end{align}
\begin{widetext}
\begin{align}
\frac{\kappa^2\p_s \zeta_{\kappa}}{D_\kappa} &= \int_{\hat q}   \frac{\Gk}{4 {\hat q}^4 h_\nu^6}
\Bigg\{
    h_D h_\nu \Big[ 
        h_\nu {\hat q}^2 \left[5 s_\nu h_D^\prime
        -6 {\hat q}^2 (r_D^{\prime\prime} s_\nu+r_\nu^{\prime\prime} s_D) + h_\nu^\prime s_D\right]
        +2 {\hat q}^4 h_\nu^\prime (10 s_\nu h_D^\prime + 7s_D h_\nu^\prime) + h_\nu^2 s_D 
    \Big]
    \nonumber \\
    &-2 h_\nu^2 {\hat q}^2 s_D \left[h_D^\prime (h_\nu+6 {\hat q}^2 h_\nu^\prime) - 2{\hat q}^2 h_\nu r_D^{\prime\prime}\right]
    - h_D^2 s_\nu \Big[ 
        3 h_\nu^2+5 h_\nu {\hat q}^2 (h_\nu^\prime - 2 {\hat q}^2 r_\nu^{\prime\prime})+25 {\hat q}^4 (h_\nu^\prime)^2
    \Big]
\Bigg\}\,,\label{eq:dsDd}
\end{align}
\begin{align}
\frac{\kappa^2\p_s \tau_\kappa}{D_\kappa} &=
\int_{\hat q} \frac{\Gk}{12 h_\nu^6 {\hat q}^4}
\Bigg\{
h_\nu^3 \left[
    \hat{q}^2 \left(
        6 h_D^\prime s_\nu+3 h_\nu^\prime s_D+4 \hat{q}^4 (r_D^{(3)} s_\nu-r_\nu^{(3)} s_D)-6 \hat{q}^2 (r_D^{\prime\prime}s_\nu+2 r_\nu^{\prime\prime} s_D)
   \right)
   -9 h_D s_\nu
\right]
\nonumber \\
&+6 h_\nu^2 \hat{q}^2 \left[
        2 h_\nu^\prime \left(
            s_\nu(h_D^\prime\hat{q}^2-h_D)+\hat{q}^4 (r_\nu^{\prime\prime} s_D-2 r_D^{\prime\prime} s_\nu)
    \right)
    +4 \hat{q}^2 r_\nu^{\prime\prime} s_\nu (h_D-h_D^\prime \hat{q}^2)+3(h_\nu^\prime)^2 \hat{q}^2 s_D
\right]
\nonumber \\
&+3 h_\nu h_\nu^\prime \hat{q}^4
   \left[
       h_\nu^\prime s_\nu (22 h_D^\prime \hat{q}^2-13 h_D)+12 h_D \hat{q}^2 r_\nu^{\prime\prime} s_\nu-2 (h_\nu^\prime)^2 \hat{q}^2 s_D
   \right]
   -60 h_D (h_\nu^\prime)^3 \hat{q}^6 s_\nu+3h_\nu^4 s_D
\Bigg\}\,,\label{eq:dstau}
\end{align}
\end{widetext}
where $s_\nu({\hat q}^2) \equiv \p_s R_\kappa^\nu({\hat q}^2) / \nu_\kappa$,
$s_D({\hat q}^2) \equiv \p_s R_\kappa^D({\hat q}^2) / D_\kappa$,
$h_\nu({\hat q}^2) \equiv \Fk+\Tk {\hat q}^2+r_\nu({\hat q}^2)$,
$h_D({\hat q}^2) \equiv 1+\Ck {\hat q}^2+r_D({\hat q}^2)$ are all dimensionless functions and their arguments are omitted for conciseness. We  used the shorthand notation
$\int_{\hat q}\equiv \int_{-\infty}^\infty \frac{d \hat q}{2\pi}$. With the choice of regulator (\ref{eq:regLPAalpha}), one has $s_\nu(x)=\alpha s_D(x) = -\alpha(\eta^D_\kappa r(x)+2x\,r'(x))$. 

\subsection{Limit of sharp cutoff and comparison with the Wilsonian RG}

Within the Wilsonian RG of Ref.~\cite{Ueno2005}, the flows of the running parameters are defined in a similar way as \eqref{eq:defparam}. However, taking momentum-derivatives is a very ambiguous operation within  Wilsonian RG with a sharp cutoff. This can be seen from a typical flow equation, {\it e.g.} Eqs. (8) and (9) of Ref.~\cite{Ueno2005}. They involve in the rhs a product of propagators at different momentum configurations, such as
\begin{equation}
  \int_{\Lambda(l)\leq |q|\leq \Lambda_0}\frac{dq}{2\pi}\int_{-\infty}^\infty \frac{d\Omega}{2\pi} |G_0(q,\Omega)|^2 G_0(p-q,\omega-\Omega) {\cal F}(p,q)
  \label{sharpcutofint}
\end{equation}
where $G_0(q,\omega)=(-i\omega+\nu q^2 +\tau q^4)^{-1}$ is the bare propagator, and ${\cal F}(p,q)$ is a simple polynomial  of momenta and of noise amplitudes $D_{\text{SKS}}$, $\zeta$, and the integration is performed over an infinitesimal momentum shell $\Lambda(l)=\Lambda_0(1-\delta l)$.
Taking derivatives of this integral with respect to $p$ may be ill-defined, as  the momentum shell introduces non-analyticities, which generate divergences 
after a specific number of differentiations (see, for example, \cite{Morris:1995af}). 

Actually, the problem is much more severe. In Eq.~(\ref{sharpcutofint}),  the cutoff is applied only to the propagators containing the momentum $q$, but not to the one containing the momentum $p-q$. This creates a problematic ambiguity, since the expression is not invariant under possible different assignments of momenta to the various propagators. This second problem can, in principle, be resolved by constructing an Exact Renormalization Group in the presence of a sharp cutoff \cite{Wegner:1972ih}, but the usual practice is to ignore this difficulty and use ad-hoc prescriptions regarding the way to cutoff various integrals.
This is a well-known problem of using a sharp cutoff to separate the fluctuation modes. This is discussed, for example, in Ref.~\cite{Morris:1995af}.

It is important to note that the use of non-smooth regulators is not merely a formal problem due to the resulting ambiguities when expanding the RG equations in momenta. For example, even when limited to approximations such as the Local Potential Approximation, in which it is not necessary to expand in momenta, it was shown that the sharp cutoff leads to poor-quality results when compared to results obtained with a smooth cutoff 
\cite{Morris:1999ba,Morris:2000hm,Litim:2000ci,DePolsi:2022wyb}.

To avoid the non-analyticities induced by the sharp cutoff, the solution is to replace it with a sufficiently smooth one, as  done in the FRG formalism.
The Wilsonian RG flow equations  can be reconstructed from FRG equations by choosing the regulator $r(\hat q^2)$ in (\ref{eq:regLPAalpha}) in the form of a sharp cutoff, namely
\begin{equation}
r(x)=\begin{cases}
+\infty,& 0 \leq x \leq 1,
\\
0,& x>1.
\end{cases}\label{eq:regSharp}
\end{equation}
However, it is clear that  derivatives of such a regulator are not defined.

For the KS equation, the flow equation (\ref{eq:dsD}) does not contain $r'$, and in (\ref{eq:dsnu}) it can be eliminated by integrating by parts, provided that the regulators are of the form (\ref{eq:regLPAalpha}).
In this case, one can safely take the sharp cutoff limit.
After integration over momentum $\hat q$, one obtains (going back to the dimensionful variables)
\begin{align}
\p_s D_\kappa &= \frac{\lambda^2(D_\kappa+\kappa^2 \zeta_{\kappa})^2}{4\pi\kappa(\nu_\kappa+\kappa^2 \tau_\kappa)^3} \,, \label{eq:dsDsharp}
\\
\p_s \nu_\kappa &= -\frac{\lambda^2(\alpha D_\kappa -\alpha\kappa^2 \zeta_{\kappa}+\nu_\kappa+3\kappa^2 \tau_\kappa)}{8\pi\alpha\kappa(\nu_\kappa+\kappa^2 \tau_\kappa)^2} \,. \label{eq:dsnusharp}
\end{align}
The flow equation for $D_\kappa$ (\ref{eq:dsDsharp}) perfectly agrees with Eq.~(13) of Ref.~\cite{Ueno2005} as expected since the definition of this flow is unambiguous even in the sharp cutoff Wilsonian flow. Note that (\ref{eq:dsDsharp}) does not depend on $\alpha$. However, the flow equation for $\nu$ (\ref{eq:dsnusharp}) differs from Eq.~(10) of Ref.~\cite{Ueno2005}, whichever $\alpha$ one chooses. We suspect an effect  of the ambiguities arising within the Wilsonian formulation, which  are absent in the FRG derivation.

Let us now comment on the flow equations for $\zeta_\kappa$ (\ref{eq:dsDd}) and $\tau_\kappa$ (\ref{eq:dstau}). They  contain the second derivative of $r$, which cannot be reduced to $r$ only by integration by parts, there remain first derivatives $r'$ that cannot be eliminated. [Note that (\ref{eq:dstau}) also contains third derivatives of the regulator, but the combination $(r_D^{(3)} s_\nu - r_\nu^{(3)} s_D)$ vanishes for the choice (\ref{eq:regLPAalpha})]. Thus,  the flow equations for $\zeta_\kappa$ and $\tau_\kappa$ do not exist in the  sharp cutoff limit. This shows that Wilsonian RG with a sharp cutoff cannot be used to tackle the KS equation. Let us emphasize once more that these issues related to the sharp cutoff are  well-documented  in the field-theoretical literature, but seem to have been overlooked in many statistical physics applications (other errors will be reported elsewhere). Therefore, we would like to convey the message that one has to be very careful when using the sharp cutoff and taking derivatives. At best, it leads to ambiguities, which can be lifted if carefully identified, but most often it leads to intrinsic divergencies that cannot be cured.

\subsection{Results within the simple DE approximation}

The flow equations \eqref{eq:flowG}--\eqref{eq:flowT}, \eqref{eq:dsD}--\eqref{eq:dstau} can  be integrated  with a suitable choice of the regulator which is not the sharp cutoff. A simple choice, which  allows one to perform integration over momentum analytically, is the (modified) Litim regulator \cite{Litim2001}
\begin{equation}
r(x) = \beta(\gamma/x-1)\theta(\gamma-x) \,, \label{eq:regLitim}
\end{equation}
where $\theta(x)$ is the Heavyside step function.
The choice of the parameters $\beta$ and $\gamma$ is discussed later.  The first derivative of this regulator is well-defined, but the second is not. However, as mentioned in the previous section, the flow equations (\ref{eq:dsDd}), (\ref{eq:dstau}) can be expressed in terms of  $r'$ and $r$ only by integration by parts, see Appendix~\ref{app:integration-by-parts}. Note that Litim regulator can  also become problematic, as the sharp cutoff, at 
high  orders of the DE, as explained, for example, in \cite{Balog:2019rrg}. However, at the present level of approximation, only its first derivative is needed and is  well defined.

With the modified Litim regulator~\eqref{eq:regLitim}, the integrals over $\hat q$ (the loop momentum) can be expressed via the Appell hypergeometric functions or performed numerically. The  flow equations \eqref{eq:flowG}--\eqref{eq:flowT}, \eqref{eq:dsD}--\eqref{eq:dstau}, which are simple ordinary differential equations, can then be  solved numerically. We use the initial condition  $\hat G_\Lambda=1, \hat F_\Lambda=-0.25, \hat C_\Lambda=0, \hat T_\Lambda=1$,  $\alpha=1$. The parameters $\beta$ and $\gamma$ of the regulators are introduced and adjusted to properly regulate the pole appearing in the flow equations at negative $\Fk$. We here choose $\beta=2$ and $\gamma=20$.
 The role of the pole is further discussed at the end of the section.

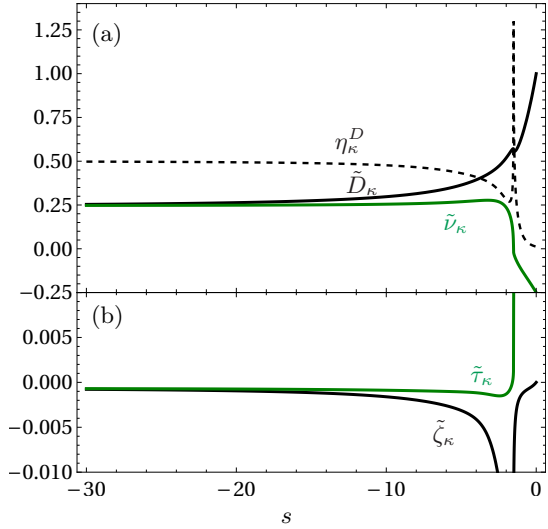
\begin{figure}
\begin{overpic}[percent,width=0.4\textwidth]{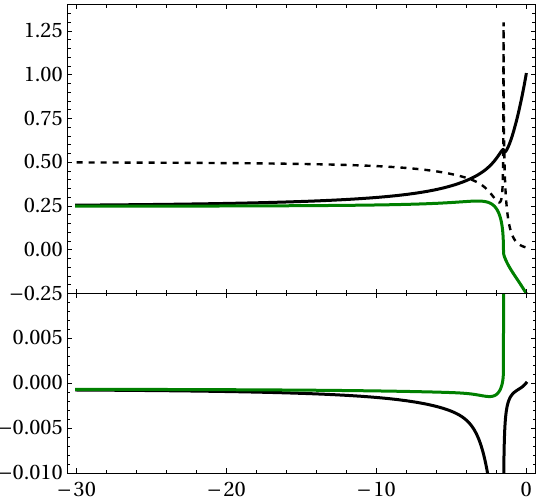}
    \put(15,85){(a)}
    \put(62,57){\textcolor{Black}{$\tilde D_\kappa$}}
    \put(80,50){\textcolor{ForestGreen}{$\tilde \nu_\kappa$}}
    \put(60,65){\textcolor{Black}{$\eta^D_\kappa$}}
    \put(15,33){(b)}
    \put(85,22){\textcolor{ForestGreen}{$\tilde \tau_\kappa$}}
    \put(50,-4){$s$}
    \put(78,11){\textcolor{Black}{$\tilde \zeta_{\kappa}$}}
\end{overpic}
  \caption{Flow of the parameters (a) $\tilde D_\kappa$,  $\tilde \nu_\kappa$ and $\eta^D_\kappa$, (b) $\tilde \zeta_{\kappa}$ and $\tilde \tau_\kappa$ defined in Eqs.~(\ref{eq:etaD}), (\ref{eq:tilde})
  as functions of the RG time $s=\ln (\kappa/\Lambda)$ calculated within the simple DE approximation. The flow starts from the KS initial condition, and reaches in the IR the KPZ fixed point, with emergent time-reversal symmetry and KPZ scaling exponents.}
  \label{fig:UenoLPA}
\end{figure}

Let us first present the results of the numerical integration of the flow, which
 is shown in Fig.~\ref{fig:UenoLPA}. To ease the comparison with Ref.~\cite{Ueno2005}, we plot dimensionful variables compensated by their fixed point scaling
\begin{align}
    \tilde D_\kappa &\equiv \kappa^{\eta_*}D_\kappa\,,\;
    &\tilde \nu_\kappa &\equiv \kappa^{\eta_*}\nu_\kappa\,,
    \nonumber \\
    \tilde \zeta_{\kappa} &\equiv \kappa^{2+\eta_*}\zeta_{\kappa}\,,\;
    &\tilde \tau_\kappa &\equiv \kappa^{2+\eta_*}\tau_\kappa\,, 
\label{eq:tilde}
\end{align}
where $\eta_*=1/2$ is the fixed point value of the anomalous dimension.
We observe that the viscosity  crosses zero and becomes positive. Its fixed point value $\tilde \nu_*$ is equal to the fixed point value $\tilde D_*$. Similarly, the parameters $\tilde\tau_\kappa$ and $\tilde \zeta_{\kappa} $  tend to the same small negative value at the fixed point ($\hat T_*,\hat C_ *\approx -0.003$). Thus, one finds at the fixed point $\hat F_* / \hat D_ * = \hat T_* / \hat C_ * = \alpha=1$, which shows that the time-reversal symmetry is emergent. Moreover, the value $\eta_*=1/2$ yields
 the critical exponents $\chi = \eta_*=1/2$ and $z=2-\eta_*=3/2$, which identify with the KPZ critical exponents. This demonstrates that the  flow for the stochastic KS equation reaches the KPZ fixed point. We checked that choosing $\alpha=1/2,\beta=2$ leads to a similar  picture, with $\hat F_* / \hat D_ * \approx \hat T_* / \hat C_ * \approx\alpha$, as expected.

Let us further comment on the pole.  Our choice of regulator parameters permits us to properly regularize the flow and avoid the singularity down to $\hat F_\Lambda=-0.25$. However, we did not manage to adjust them to regularize the flow at $\hat F_\Lambda=-1$. This is due to  a rapid decrease of $\Tk$ together with a very slow increase of $\Fk$ towards positive values, which causes the flow to cross the pole. Let us emphasize that the presence of a pole is intrinsic to the KS dynamics. This is completely different from the divergencies encountered when taking the sharp cutoff limit, which are a mere artifact of choosing a non-analytic regulator, and are hence not physical.

The fact that negative coefficients  $\Fk$ and $\Tk$  in the expansion around zero momentum destabilize the flow signals that this point is not a good expansion point, and that the momentum dependence of the effective viscosity is important and cannot be neglected.
We show in Sec.~\ref{sec:NLO} the results for a functional approximation.
Indeed, the full momentum- and frequency-dependent functions $\hat f^\nu_\kappa(\hat\omega,\hat p)$ and $\hat f^D_\kappa(\hat\omega,\hat p)$ (defined in the next section) reach the same (concave) shape at the fixed point (see Sec.~\ref{sec:NLO}) which is consistent with the negative and equal values found for $\Ck$ and $\Tk$. In the functional setting, the flow adjusts itself to overcome the divergence, even at $\hat F_\Lambda=-1$.
From the shape 
of the flowing functions $\hat f^{\nu,D}_\kappa(\hat\omega,\hat p)$
one can infer that if a momentum expansion is performed, it should be done around a non-zero $p_0$, corresponding to the minimum of the denominator of the regulated propagator. This is explained and implemented in Sec.~\ref{sec:noise} and Appendix~\ref{app:p0-LPA}.

\section{The NLO approximation}
\label{sec:NLO}

\subsection{NLO ansatz}

We now turn to a more expressive   approximation, termed the next-to-leading order (NLO) approximation~\cite{Kloss2012}. The idea behind this approximation is to devise an ansatz for $\Gamma_\kappa$ which preserves  the symmetries of the KPZ equation and accurately describes momentum- and frequency-dependence of the 2-point vertex functions. The dependence of $\Gamma_\kappa$ on the fields is truncated to the minimal order. Hence, this approximation allows for momentum- and frequency-dependent effective noise and viscosity. The NLO ansatz is given by
\begin{align}
\Gamma_{\kappa}[\varphi,\bvarphi]= \int_{t,x} &\bvarphi \Big\{
    \p_t \varphi - f_\kappa^\nu(\p_t,\p_x)\p_x^2\varphi\nonumber\\
     &-\frac{\lambda}{2}(\p_x\varphi)^2 -f^D_\kappa(\p_t,\p_x)\bvarphi
\Big\} \,.
\label{eq:GammakNLO}
\end{align}
With this ansatz, the 1-PI two-point functions take the form in Fourier space
\begin{align}
\Gamma^{(1,1)}_{\kappa}(\omega,p) &= i \omega + p^2 f^\nu_\kappa(\omega,p)\,,
\nonumber \\
\Gamma^{(0,2)}_{\kappa}(\omega,p) &=-2f^D_\kappa(\omega,p)\,,
\label{eq:Gamma2NLO}
\end{align}
and the effective noise and viscosity are defined as $\nu_\kappa=f^\nu_\kappa(0,0)$, $D_\kappa=f^D_\kappa(0,0)$.
The time-reversal symmetry constraint  now writes $f^\nu_\kappa=\alpha f^D_\kappa$  according to (\ref{eq:FDT}),(\ref{eq:FDT1}).

Note that we did not promote  $\lambda$ to a function  $f^\lambda_\kappa(\omega,p)$ because the Galilean invariance imposes
\begin{equation}
f^\lambda_\kappa(\omega,p=0)=\lambda\,,
\label{eq:f_lambda_p0}
\end{equation}
and furthermore, in 1D,  the time-reversal symmetry of the KPZ equation imposes within the NLO approximation~\cite{Canet2011,Kloss2012}
\begin{equation}
f^\lambda_\kappa(\omega,p)\equiv\lambda\,.
\label{eq:f_lambda}
\end{equation}
In the KS context, only the constraint \eqref{eq:f_lambda_p0} holds at the microscopic scale, but we choose as a simplification 
 to fix the full constraint \eqref{eq:f_lambda}. The reason is that it was shown  in  Ref.~\cite{Kloss2012} that the function $f^\lambda_\kappa$ has little influence and only leads to a small quantitative change, we thus simply neglect it.
The KS initial condition then corresponds to
\begin{equation}
 f^\nu_\Lambda(\omega,p) = \nu_\Lambda  + \tau_\Lambda p^2\,, \quad  f^D_\Lambda(\omega,p) = D_\Lambda\, ,\quad \nu_\Lambda <0\,.
\end{equation}
Note that since we are considering full momentum dependent functions, there is no need to introduce a parameter such as $\zeta_\kappa$. The relevant momentum content is simply generated by the flow.

One can derive the flow equations for $f^{\nu,D}_\kappa$ as
\begin{align}
&\p_s f^\nu_\kappa(\omega,p) = \frac{1}{p^2}\p_s \Gamma^{(1,1)}_\kappa(\omega,p) \,,
\nonumber \\
&\p_s f^D_\kappa(\omega,p) = -\frac{1}{2}\p_s\Gamma^{(0,2)}_\kappa(\omega,p)\,.
\end{align}
The rhs of these equations are  integrals over $\omega',p'$ whose integrands contain $f^{\nu,D}_\kappa(\omega',p')$. In the NLO approximation, the $\omega'$-dependence of $f^{\nu,D}_\kappa$ is neglected in the integrands: $f^{\nu,D}_\kappa(\omega',p') \equiv f^{\nu,D}_\kappa(0,p')$. This permits one to perform the $\omega'$-integration analytically. The detailed derivation of the NLO equations can be found in~\cite{Kloss2012}. 
 We also refer the interested reader to this reference for a  discussion of the NLO approximation in the KPZ context.

\subsection{Dimensionless flow equations}

In order to study fixed points and the corresponding scaling regime, one  needs to define dimensionless quantities. The usual choice for KPZ is
\begin{align}
&\hat f^{\nu}_\kappa(\hat\omega,\hat p) = f^\nu_\kappa(\omega,p) / \nu_\kappa
\,,\,
\hat f^{D}_\kappa(\hat\omega,\hat p) = \hat f^D_\kappa(\omega,p)/D_\kappa
\,,\nonumber\\
&\hat p=\hat p
\,,\,
\hat\omega=\omega/(\kappa^2\nu_\kappa)
\,,\,
\hat g_\kappa = \kappa^{-1} \lambda^2 D_\kappa / \nu_\kappa^3\,.
\label{eq:hatg}
\end{align}
Again, in the KS context,  one encounters the problem of vanishing $\nu_\kappa$. One can make the same choice as for the simple DE approximation. However, we found that, within the NLO approximation,  another choice leads to flow equations which are easier to numerically integrate. This choice is the following:
\begin{align}
&\hat f^D_\kappa(\hat\omega,\hat p) \equiv f^D_\kappa(\omega,p) / D_\kappa\,,
\nonumber\\
&\hat f^\nu_\kappa(\hat\omega,\hat p) \equiv f^\nu_\kappa(\omega,p) / A_\kappa\,,\,
A_\kappa \equiv |\nu_\Lambda|\kappa^{-\eta^A}\,,
\end{align}
where the dimensionless momentum and frequency are \begin{align}\hat p \equiv p/\kappa\,,
\quad
\hat \omega \equiv \omega/(\kappa^2 A_\kappa)\,,
\label{eq:dimlessGrid}
\end{align}
and $\eta^A$ is fixed to $\eta_*=1/2$. 
We  introduce the dimensionless coupling as
\begin{align}
\hat g^A_\kappa = \kappa^{-1} \lambda^2 D_\kappa / A_\kappa^3\,,
\label{eq:hatgA}
\end{align}
and  choose the regulators accordingly:
\begin{align}
&R_\kappa^D(p^2) = D_\kappa r_D(\hat p^2)\,,\,R_\kappa^\nu(p^2) = A_\kappa r_\nu(\hat p^2)\,,
\nonumber\\
&r_\nu=C_\nu r_D\equiv C_\nu r\,.
\label{eq:regFlowA}
\end{align}
The role of the constant $C_\nu$, which allows for a smooth continuation of the {\it flow-A} to a flow with standard KPZ adimensionalization, is discussed in Appendix~\ref{app:flowA}.
The dimensionless equations are given by
\begin{align}
\p_s \hat g^A_\kappa &= \hat g_\kappa(3\eta^A-\eta^D_\kappa-1)\,,
\nonumber\\
\p_s\hat f^D_\kappa(\hat\omega,\hat p) &= 
\big(\eta^D_\kappa + \hat p \p_{\hat p}+(2-\eta^A)\hat \omega \p_{\hat \omega}\big) \hat f^D_\kappa(\hat\omega,\hat p) 
\nonumber\\
&+ \hat I^D_\kappa(\hat\omega,\hat p)\,,
\nonumber\\
\p_s\hat f^\nu_\kappa(\hat\omega,\hat p) &=  \big(\eta^A + \hat p \p_{\hat p}+(2-\eta^A)\hat \omega \p_{\hat \omega}\big)\hat f^\nu_\kappa(\hat\omega,\hat p) 
\nonumber\\
&+ \hat I^\nu_\kappa(\hat\omega,\hat p)\,,
\label{eq:flowAdimlessEqs}
\end{align}
where the nonlinear contributions $\hat I^D_\kappa$ and $\hat I^\nu_\kappa$,  defined as
$\hat I^D_\kappa(\hat\omega,\hat p) \equiv \p_s f^D_\kappa(\omega,p)/D_\kappa$,
 and $\hat I^\nu_\kappa(\hat\omega,\hat p) \equiv \p_s f^\nu_\kappa(\omega,p)/A_\kappa$, are given by
\begin{align}
&\hat I^\nu_\kappa(\hat\omega,\hat p) = 
\int_{\hat q, \hat \omega'} \frac{
    2 \hat g^A_\kappa \hat q^2 (\hat p+\hat q)
}
{
    \hat p P_\kappa(\hat\omega',\hat q)^2 P_\kappa(\hat\omega+\hat\omega',\hat p+\hat q)
}\nonumber\\
&\times\Big[
    \hat q (\hat p+\hat q) s_\nu(\hat q^2)h^D_\kappa(0,\hat p+\hat q) 
    \Big(
        P(\hat\omega',\hat q)-2\hat\omega'^2
    \Big) \nonumber\\
    &+(\hat p+\hat q)^2 h^\nu_\kappa(0,\hat p+\hat q) \Big(
        s_D(\hat q^2) P(\hat\omega',\hat q) 
        \nonumber\\
        &- 2 \hat q^4 s_\nu(\hat q^2) h^D_\kappa(0,\hat q) h^\nu_\kappa(0,\hat q)
    \Big)
\Big]
\,,\\
&\hat I^D_\kappa(\hat\omega,\hat p) = \int_{\hat q, \hat \omega'} \frac{
    2 \hat g^A_\kappa \hat q^2 (\hat p+\hat q)^2 h^D_\kappa(0,\hat p+\hat q)
}
{
    P_\kappa(\hat\omega',\hat q)^2 P_\kappa(\hat\omega+\hat\omega',\hat p+\hat q)
}\nonumber\\
&\times\Big[
    s_D(\hat q^2) P_\kappa(\hat\omega',\hat q) - 2\hat q^4 s_\nu(\hat q^2) h^D_\kappa(0,\hat q) h^\nu_\kappa(0,\hat q)
\Big]\,,
\end{align}
where, consistently with notation of \eqref{eq:dstau},
$s_\nu({\hat q}^2) \equiv \p_s R_\kappa^\nu({\hat q}^2) / A_\kappa$,
$s_D({\hat q}^2) \equiv \p_s R_\kappa^D({\hat q}^2) / D_\kappa$,
$h_{\nu,D}(\hat\omega,\hat q) \equiv \hat f^{\nu,D}_\kappa(\hat\omega,\hat q)+r_{\nu,D}({\hat q}^2)$, and
$P_\kappa(\hat\omega,\hat p) \equiv  \hat\omega^2+[\hat p^2 h_\nu(0, \hat p)]^2$. We use the shorthand notation $\int_{\hat\omega,\hat q}=\int_{-\infty}^\infty \frac{d\hat\omega}{2\pi}\int_{-\infty}^\infty \frac{d\hat q}{2\pi}$ as previously.
Let us call this normalization scheme \textit{flow-A}.
As we will show below,  the flow equations (\ref{eq:flowAdimlessEqs}) have an IR fixed point which has the KPZ scaling. Let us emphasize that this scaling is not forced by our choice $\eta^A=1/2$, as discussed in Appendix \ref{app:flowA}. Moreover, we checked that adimensionalizing with $D_\kappa$ leads to qualitatively similar results, although more difficult to numerically stabilize at large negative $\nu_\Lambda$.

\subsection{Numerical integration: two-grid scheme}

The NLO approximation proved to be efficient in capturing both UV and IR scaling regimes of the KPZ equation when solving the flow equations within a special numerical scheme that we call the \textit{two-grids} approach \cite{Benitez2009, Mathey2017, Gosteva2025KPZtwogrids, Paris2026}. The idea is as follows. The NLO approximation was devised to provide an accurate description of the IR properties, but its reliable applicability for the UV region is not granted. 
From a numerical point of view, this is reflected in the fact that the flow equations (\ref{eq:flowAdimlessEqs}) are defined on the dimensionless grid (\ref{eq:dimlessGrid}), which spans a shrinking area in the dimensionful $(\omega,p)$-space, as $\kappa$ changes from $\Lambda$ to $0$. The NLO equations can be extended also to the regime of {\it fixed} momenta and frequencies, to yield an approximate flow in this regime. The interesting point is that, in the large-momentum regime $|p|\gg\kappa$, one can obtain a closed set of equations {\it without} further approximation than the large $|p|$ limit.  In order to do this, on can derive a second set of equations, which we can call the \textit{large-$p$} equations, that become exact  in the UV limit \cite{Fontaine2023Unpredicted, Gosteva2024Burgers}. The derivation (similar to the one for the Navier-Stokes equation in the work \cite{CanetPRE2016}, where it was obtained for the first time) is based on the symmetries of the model and on the assumption that the external momentum $p$ is large: $|p|\gg\kappa$. The \textit{large-$p$} equations have a simple form
\begin{align}
&\p_s C_{\kappa}(\varpi, p) =  {p^2}
\int_{\omega} \dfrac{C_{\kappa}(\omega+\varpi,  p) -
C_{\kappa}(\varpi,  p)}{\omega^2}\,  {\cal J}_\kappa(\omega)\,,
\nonumber\\ \label{eq:large-p}
&{\cal J}_\kappa(\omega)\equiv  \int_{q} q^2\, \tilde{\p}_sC_{\kappa}(\omega, q)\,,
\end{align}
where the operator $\tilde{\p}_s$ only acts on the $s$-dependence of the regulators, and where $C_\kappa$ is the 2-point connected correlation function defined in \eqref{eq:corr}. A  closed form of the equation was also obtained for higher-order correlation functions~\cite{Tarpin2018}.

In the \textit{two-grids} scheme, we use a dimensionless grid $\{(\hat\omega_{m},\hat p_{n})\}$
and a dimensionful grid $\{(\omega_i,p_j)\}$ to solve numerically the \textit{small-p} (NLO) equations (\ref{eq:flowAdimlessEqs}) and the \textit{large-p} equation (\ref{eq:large-p}) coupled together in the following way.
We solve the \textit{small-p} equations, as usual, on the dimensionless grid (see, for example, \cite{Kloss2012}). At $\kappa=\kappa_j$, when a point $p_j$ in the dimensionful grid exits\footnote{
    The exit criterion is $p_j\geq \kappa \hat p_{\text{exit}}$ where $\hat p_{\text{exit}}$ is chosen as the momentum value  at which the regulator  becomes negligible, typically $4/\sqrt{\beta}$ ({\it eg} $\hat p_{\text{exit}}\simeq 5.66$ for $\beta=0.5$). Technically, $\hat p_{\text{exit}}$ then coincides with the limit of numerical $\hat q$-integration,  $\hat q_{\text{max}}$. The value of $\hat p_{\text{exit}}$ is therefore not strictly prescribed. We checked that, indeed, varying it around $\hat q_{\text{max}}$ does not change the result for $C(\omega,p)$.}
the applicability region of the \textit{small-p} equations, we record the flow parameters. They are used to construct an initial condition (IC) for the \textit{large-p} equation at $p=p_j$ and $\forall\omega_i$ which is solved on the dimensionful grid. The procedure is described in detail in \cite{Gosteva2025KPZtwogrids}.

It was noticed in \cite{Gosteva2025KPZtwogrids} that the result of the integration of the \textit{large-p} equation does not lead to significant difference compared to the IC. In other words, the NLO result is an approximate fixed-point solution for the \textit{large-p} equation. It shows that, in fact, the NLO approximation is capable to describe a wider range of momenta in UV than one would expect. In this work, we simply record the IC on the dimensionful grid instead of solving the \textit{large-p} equations.
The last simplification that one can perform is to replace the NLO approximation with the LO (leading order) approximation~\cite{Canet2010}, where only  the momentum dependence of $\hat f^{D,\nu}_\kappa(\omega,p)$ is kept, while the frequency dependence is neglected. 
We will resort to it  in Sec.~\ref{sec:noise}.

Thus we can use the full potential of the NLO flow equations and,  by recording their output at the right scale, retrieve from them both IR and UV properties in a wide range of dimensionful momenta and frequencies. For the KPZ equation, the \textit{two-grids} scheme was successfully used to capture both EW or IB (in UV) and KPZ (in IR) scaling regimes~\cite{Gosteva2025KPZtwogrids}. It was also applied to the  stochastic Navier-Stokes equation and predicted two scaling regimes, in line with direct numerical simulations~\cite{Gosteva2025Euler}. For the KS equation, it showed an impressive qualitative description for the correlation function compared with large-scale  numerical simulations~\cite{KSI}.

Before presenting the \textit{two-grids} result for $C(\omega,p)$, note that an asymptotical solution of (\ref{eq:large-p}) can be found, and it reads~\cite{Tarpin2018,Fontaine2023Unpredicted}
\begin{equation}
    C(t,p) = C(0,p){\times}
    \begin{cases}
        \exp\left( - \mu_0 (pt)^2 \right),\quad t\ll \tau_c
        \\
        \exp\left( - \mu_{\infty} p^2 |t| \right),\quad t\gg \tau_c
    \end{cases}
\label{eq:Casymp}
\end{equation}
for large $p$, where $\mu_0$, $\mu_{\infty}$ are non-universal constants and $\tau_c$ is a characteristic time scale. One observes that, at large momenta, that is, in the UV, there are two scaling regimes:  $z=1$ (small times, large frequencies) and $z=2$ (large times, small frequencies).

\subsection{Results within the NLO approximation}

\begin{figure}
\begin{overpic}[percent,width=0.41\textwidth]{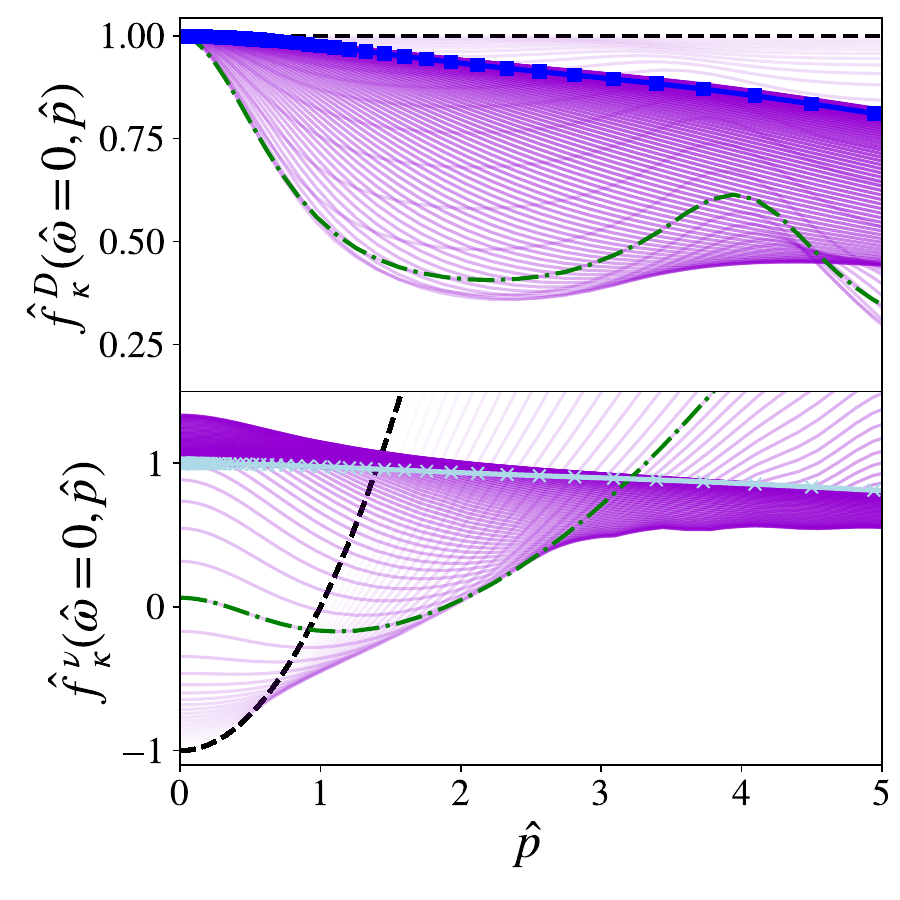}
  \put(-10,90){(a)}
  \put(-10,50){(b)}
\end{overpic}
\begin{overpic}[percent,width=0.41\textwidth]{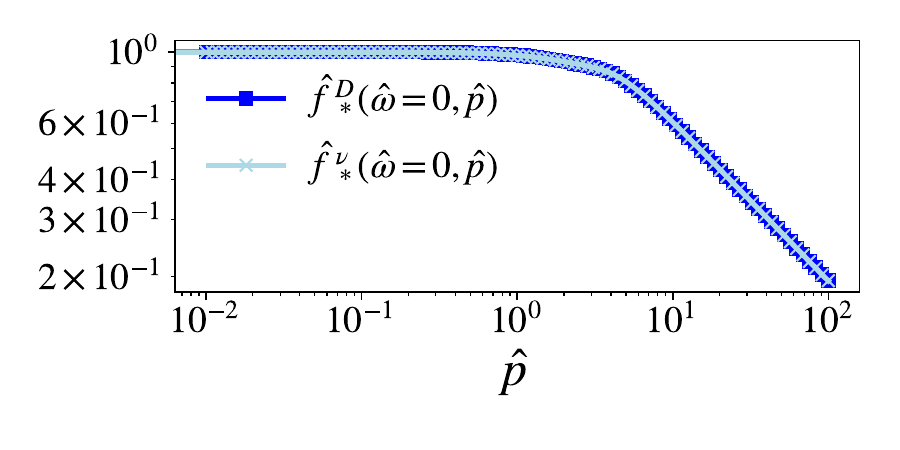}
  \put(-10,40){(c)}
\end{overpic}
  \caption{NLO flow of the functions (a) $\hat f^{D}_\kappa$ and (b) $\hat f^\nu_\kappa$ (in the \textit{flow-A} before the continuation). 
  The dashed black line marks the initial condition, darker violet indicate later RG times (larger length scales), the green dash-dotted line highlights the scale at which the effective viscosity just crosses zero, the blue lines with symbols mark the fixed point functions, which are compared in (c).}
  \label{fig:flow}
\end{figure}

We display in Fig.~\ref{fig:flow} typical evolution of the functions $\hat f^{\nu}_\kappa(\hat \omega, \hat p)$ and $\hat f^D_\kappa(\hat \omega, \hat p)$ (at $\hat\omega=0$) calculated with the \textit{flow-A} NLO approximation, and solved using the \textit{two-grids} scheme with regulators~(\ref{eq:regFlowA}), where
\begin{equation}
\hat r(\hat p^2) = \alpha\exp(-\beta \hat p^2) / \hat p^2\,,
\label{eq:regulator}
\end{equation}
$\alpha=8$, $\beta=0.5$. We checked that the Wetterich regulator $\hat r(\hat p^2) = 2 / (\exp(\hat p^2)-1)$ yields similar results. 
One observes that, starting from the KS initial condition (indicated by  black lines), the functions smoothly deform to acquire a fixed point form when $\kappa\to 0$ (indicated by thick violet lines). Fig.~\ref{fig:flow}(c) shows that the fixed point forms $\hat f^{\nu}_*$
 and $\hat f^{D}_*$ coincide, which demonstrates that time-reversal symmetry is emergent at large distances, and they match the KPZ fixed-point function. Thus, in accordance with the results from the DE approximation, we find that the IR behavior of the stochastic KS equation belongs to the KPZ universality class.

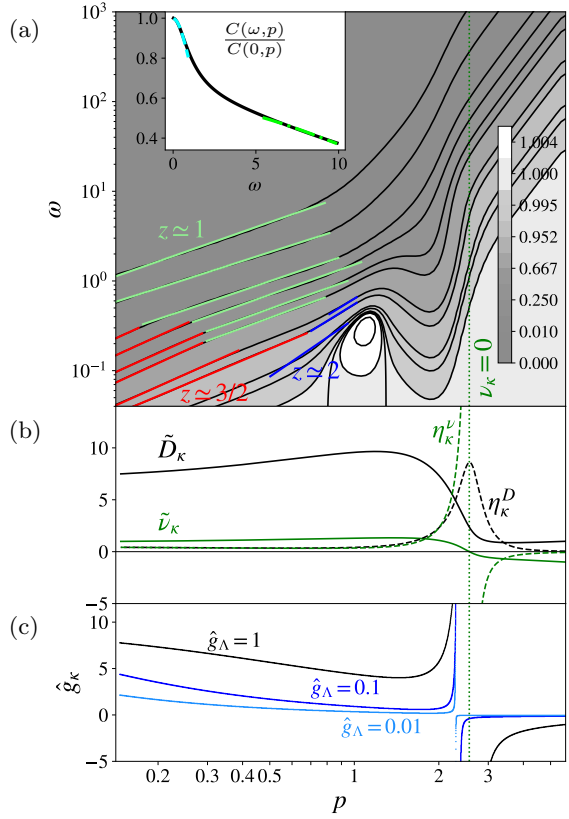
\begin{figure}
\begin{overpic}[percent,width=0.4\textwidth]{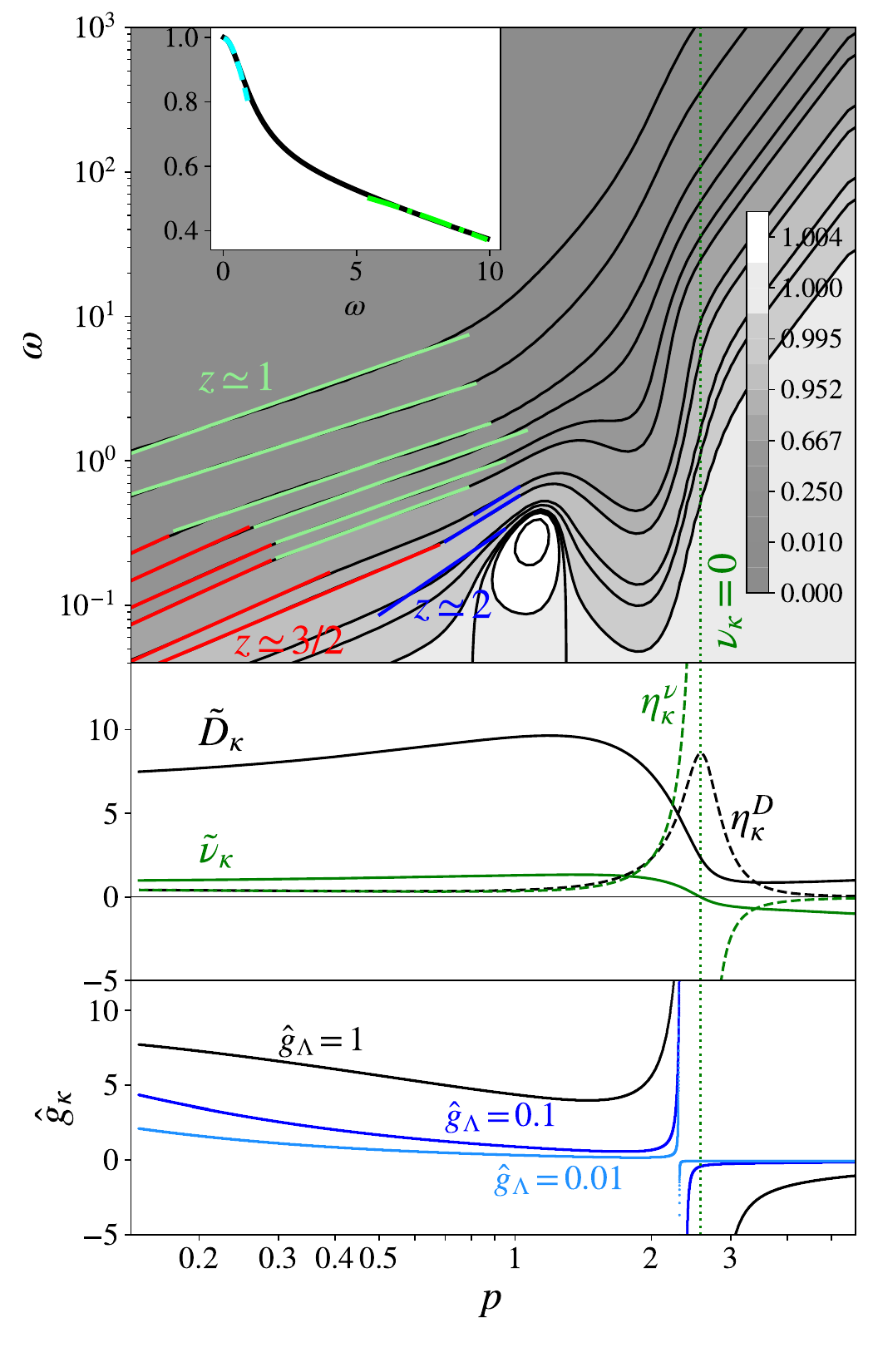}
    \put(-3,95){(a)}
    \put(22,94){$\frac{C(\omega,p)}{C(0,p)}$}
    \put(-3,48){(b)}
    \put(-3,25){(c)}
\end{overpic}
\caption{
  (a): Correlation function $C(\omega,p)/C(0,p)$ in the $(\omega,p)$ plane obtained within the {\it flow-A} NLO approximation, calculated for $\hat g_\Lambda =1$ ($\lambda=D_\Lambda=-\nu_\Lambda=1$). 
  The red, blue, and green lines  show linear fit of the contour lines. The fits yield a slope $z$ in the range $z\in[1.32, 1.44]$ for the red lines (onset of the KPZ regime),  in the range $z\in[1.94, 2.14]$ for the blue lines (EW regime), and in the range $z\in[1.01, 1.15]$ for the green lines (IB regime).
  (Inset): Fit of the frequency dependence of correlation function   at $p=2.07$ (black) close to the scale when $\nu_\kappa=0$ with the Fourier representations of the asymptotic prediction \eqref{eq:Casymp_Fourier} (cyan and green).
  (b): Effective noise $\tilde D_\kappa \equiv \kappa^{\eta_*} D_\kappa$ and viscosity $\tilde \nu_\kappa \equiv \kappa^{\eta_*}\nu_\kappa$ (black and green solid lines), and anomalous dimensions $\eta^D_\kappa$ and $\eta^\nu_\kappa$ (black and green dashed lines) as functions of the RG (momentum) scale, calculated for $\hat g_\Lambda =1$.
  (c): Effective coupling parameter $\hat g_\kappa$ (\ref{eq:hatg})
  calculated for different initial $\hat g_\Lambda$ (black: 1, blue: 0.1, lightblue: 0.01).
  In (b) and (c), the RG scale $\kappa$ is rescaled by $\hat p_{\text{exit}}=5.66$ (see footnote 3), such that it qualitatively corresponds to the momentum axis $p$. The vertical dotted line marks the momentum scale where $\nu_\kappa$ vanishes for $\hat g_\Lambda=1$.
}
\label{fig:NLO_C}
\end{figure}

From these fixed point functions, one deduces the correlation function 
\begin{equation}
C_*(\omega,p) = 
    \frac{2 f^D_*(\omega,p)}{\omega^2 + \big(p^2f^\nu_*(\omega,p)\big)^2}
\end{equation}
The result is shown in Fig.~\ref{fig:NLO_C}(a).
We identify three scaling regimes.
The KPZ scaling $z=3/2$ emerges in the small-$p$ region,  sooner for smaller $\omega$.
It is  preceded by the EW scaling $z=2$ in the larger-$p$, small-$\omega$ region, and the IB scaling $z=1$ in larger-$p$ and larger-$\omega$ region\footnote{
    The $z=4$ regime at the largest momenta, in the deep UV, is prescribed by the KS initial condition of the flow equations and is not universal. Changing the initial condition affects this region, but does not affect the KPZ and IB/EW regions.
}. The presence of the EW scaling at larger $p$'s agrees with the numerical observations of  this regime for small system size \cite{Hyman1986,Zaleski1989, Sneppen1992, Hayot1993, Ueno2005} and with the asymptotic solution (\ref{eq:Casymp}). Note that the correlation function has a maximum (light area on the colormap in Fig.~\ref{fig:NLO_C}(a)) which determines the behavior of the contour lines near it: steeper contour lines (EW regime) to the left and less steep lines (IB regime) above. The FRG thus gives the full spatio-temporal description of the correlation function, with two regimes at intermediate/large wavenumbers (IB and EW) coexisting at different frequency, while the effective coupling $\hat g_\kappa$ (the usual “observable” in the perturbative RG) would only suggest the presence of the EW regime corresponding to small $\hat g_\kappa$ before the KPZ plateau (see Fig.~\ref{fig:NLO_C}(c); note that the KPZ plateau is not present in this figure, as it is zoomed in the large-momentum region to better resolve  the transition between IB, EW and KPZ regimes).

The correlation function at large-$\omega$ is in accordance with the Fourier transform of the small-$t$ asymptotic solution (\ref{eq:Casymp}) in a whole range of momenta corresponding to the IB scaling, starting from (smaller than) the scale $\kappa=\kappa_{\nu=0}$ when the effective viscosity vanishes. A cut of $C(\omega,p)$ at $p$ corresponding to this scale is presented in the inset of Fig.~\ref{fig:NLO_C}(a). It is in the vicinity of $\kappa_{\nu=0}$ that both the small-$\omega$ and large-$\omega$  region are best modeled by the Fourier transform corresponding to both the large-$t$ and small-$t$ limit of the exact asymptotic solution (\ref{eq:Casymp}), respectively 
\begin{align}
\frac{C(\omega,p)}{C(0,p)} \simeq
\begin{cases}
    \frac{2\mu_\infty p^2}{(\mu_\infty^2p^4+\omega^2)}\,, \text{ small }\omega\,,
    \\
    \sqrt{\frac{\pi}{\mu_0 p^2}}\exp\Big(\frac{-\omega^2}{4\mu_0 p^2}\Big)\,,\text{ large }\omega\,.
\end{cases}
\label{eq:Casymp_Fourier}
\end{align}

The flow of the parameters $\nu_\kappa$, $D_\kappa$ extracted from the function $f_\kappa^\nu$, $f_\kappa^D$, together with the corresponding anomalous dimensions $\eta_\kappa^\nu$, $\eta_\kappa^D$ are shown in Fig.~\ref{fig:NLO_C}(b). One observes that, indeed, $\nu_\kappa$ crosses zero, leading to the onset of the IB scaling regime. One can also link the EW regime with the region where the effective nonlinearity $\hat g_\kappa$  remains small. The position of the IB region to the left from the $\nu_\kappa=0$ point is the same as in the KPZ equation \cite{Gosteva2025KPZtwogrids}, where the IB regime was observed 
when
$\nu_\kappa$ already departed from zero. For the EW regime the situation is similar: it persists when $\hat g_\kappa$ is no longer negligible and already approaches the value of the KPZ plateau. This further suggests that the KPZ regime requires very large scales to develop.

Fig.~\ref{fig:NLO_C}(a) is obtained for a large value of $\hat g_\Lambda=1$. Thus, the EW scaling regime is  barely visible, as well as the plateau of small $\hat g_\kappa$ which is not very  pronounced. This plateau enlarges at smaller $\hat g_\Lambda$, as shown in Fig.~\ref{fig:NLO_C}(c), where we vary $D_\Lambda$ while keeping $\lambda=-\nu_\Lambda=1$. This shows that the EW regime is more extended when the microscopic noise $D_\Lambda$ is small.  In other words, the KPZ regime starts at smaller lengthscales for the noisy KS equation than for the deterministic one, where the effective noise is produced only as the result of the chaotic nature of the dynamics, and the extent of the EW regime (corresponding to the plateau of small values of $\hat g_\kappa$) is wider.
This result is in qualitative agreement with the estimations of Ref.~\cite{Sneppen1992} and with the observations of Ref. \cite{Vercesi2024}.
We further quantify  the interplay between the EW and the KPZ regimes is  Sec.~\ref{sec:noise}.

Let us emphasize that within the NLO approximation, the flow does not diverge despite the $\nu_\Lambda=-1$ initial condition and passes through the $\nu_\kappa=0$ point smoothly (in contrast to the flow in Sec.~\ref{sec:LPA}). The negative part of the flowing function $\hat f^\nu_\kappa(\hat\omega,\hat p)$ drifts from $\hat p=0$ to larger momenta, and is still present even after $\nu_\kappa$ becomes positive (see Fig.~\ref{fig:flow}(a)). In this more expressive approximation, the flow adapts itself in a non-singular way to build from the KS initial condition the KPZ universal regime. If one expands this function around $p=0$ as in the DE approximation (which corresponds to what is done in perturbative RG), it yields a negative curvature (as witnessed by $\Fk>0$ and $\zeta_{\kappa}<0$) which is not stabilized at large $p$ since the DE is polynomial in $p$, in contrast with the NLO where the functions decay to zero at large $p$. This shows that a polynomial expansion in momentum around zero is not well adapted to capture the KS dynamics.
However, such an expansion can be much better performed, provided another expansion point is chosen, as explained in the next section. This point corresponds to the largest unstable mode (minimum of the denominator of the regulated propagator).
This is explained in the next section.

\section{Limit of vanishing noise, deterministic KS equation}
\label{sec:noise}

Let us now investigate the link between the stochastic and the deterministic KS equations. In particular, the salient question is the fate of the KPZ regime, {\it i.e.}  how the scale of the onset of the KPZ regime depends on the microscopic noise level $D_\Lambda$, and its behavior in the deterministic limit $D_\Lambda\to0$. The direct computation with $D_\Lambda=0$ corresponds to $\hat g_\Lambda=\Lambda^{-1} \lambda D_\Lambda/|\nu_{\Lambda}|^3=0$. Hence it yields the EW scaling,  leaving $\nu_{\Lambda}=-1$ unchanged since there is no RG flow in this trivial limit (the same was reported in Ref.~\cite{Ueno2005} for zero noise). Instead, we explore a wide range of small $\hat g_\Lambda$, {\it i.e.} small  $\hat D_\Lambda$, and extrapolate the result to $\hat g_\Lambda\to0$. In this parameter region, the NLO or LO approximation is difficult to stabilize numerically because the minimum of the denominator of the regulated propagator is getting very close to zero. We hence resorted to an alternative approximation, inspired by this observation, which  we call $p_0$-DE, as it consists in a momentum expansion, performed around  a non-zero $p_0$. This momentum is chosen as the momentum where the propagator denominator exhibits a minimum. 

In this approximation $\hat f^\nu_\kappa(\hat\omega,\hat p)$ is described by only three independent parameters, and $\hat f^D_\kappa(\hat\omega,\hat p)\equiv 1$ (that is, $f^D_\kappa(\omega,p)\equiv D_\kappa$) since this function does not play an important role in the KS flow. The corresponding flow equations are derived in Appendix~\ref{app:p0-LPA}.

\begin{figure}
\begin{overpic}[width=8cm, percent]{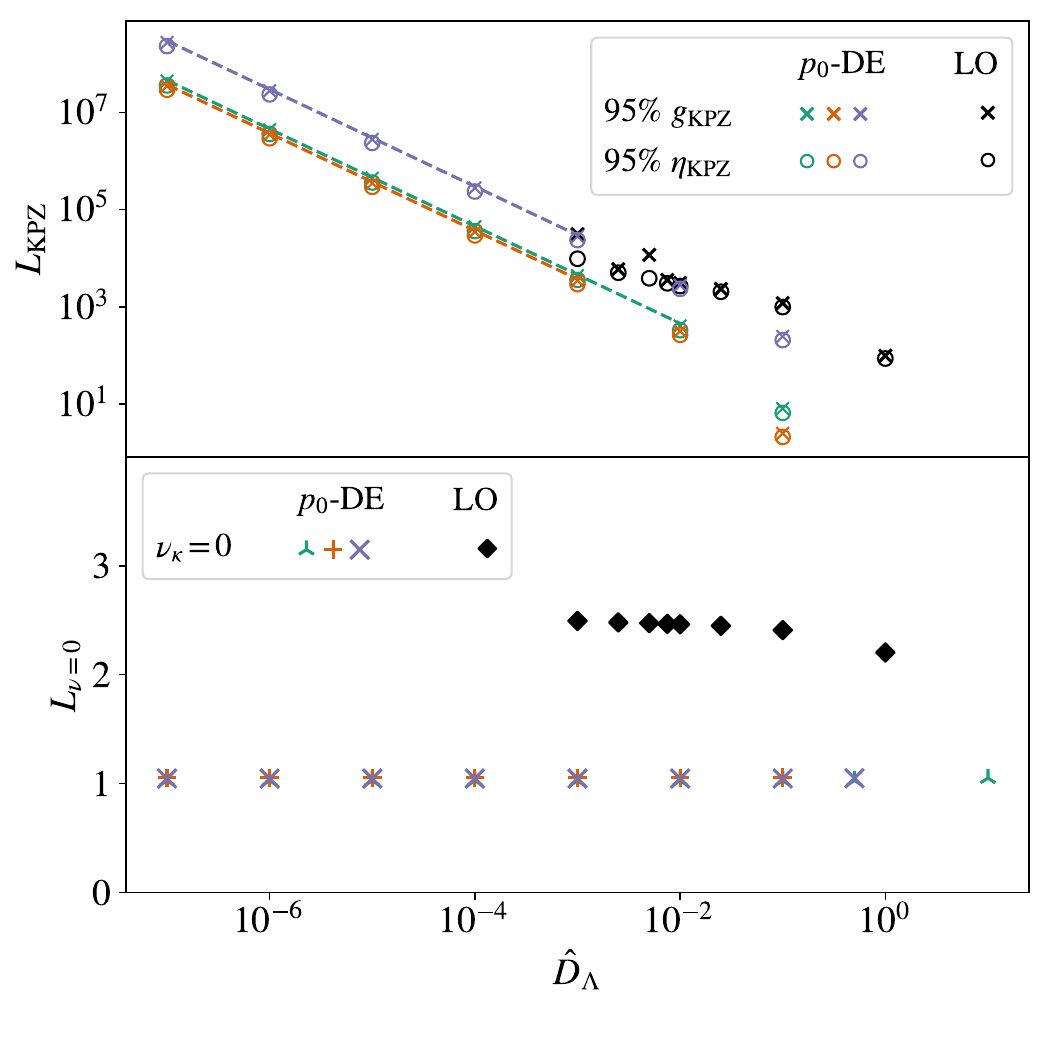}
  \put(-5,95){(a)}
  \put(-5,50){(b)}
\end{overpic}
  \caption{
  (a): Length scale at which the KPZ regime is attained, as a function of the noise amplitude $\hat D_\Lambda$. Crosses represent the scale when the effective coupling parameter $\hat g_\kappa$ reaches 95\% of its KPZ fixed point value, circles represent the same for $\eta^D_\kappa$. 
  (b): Length scale at which the effective viscosity vanishes as a function of the noise amplitude.
  For both panels, colored symbols show $p_0$-DE results with regulator (\ref{eq:regulator}) and parameters 
  $C_\nu=0.25$, $\alpha=6$, $\beta=0.5$ (green),
  $C_\nu=0.25$, $\alpha=6$, $\beta=1$ (orange),
  $C_\nu=0.5$, $\alpha=6$, $\beta=1$ (violet), while
  black symbols show LO results ($C_\nu=1$, $\alpha=8$, $\beta=0.5$).
  The dashed lines indicate a linear fit.}
  \label{fig:L_of_g}
\end{figure}

To investigate the approach to the IR regime, we decrease $D_\Lambda$ (or, equivalently, $\hat g_\Lambda$) towards the zero limit and we record two scales. The first one is the RG scale $\kappa_{\nu=0}$ and corresponding lengthscale $L_{\nu=0} \equiv \kappa_{\nu=0}^{-1}$ at which the effective viscosity $\nu_\kappa$ crosses zero, which is shown in Fig.~\ref{fig:L_of_g}(b). The second one,  shown in  Fig.~\ref{fig:L_of_g}(a), is  the RG scale $\kappa_{\text{KPZ}}$ and corresponding lengthscale $L_{\text{KPZ}} \equiv \kappa_{\text{KPZ}}^{-1}$ at which the KPZ regime emerges,
which we define as the scale where the effective coupling $\hat g_\kappa$ or anomalous dimension $\eta^D_\kappa$ reaches 95\% of its value at the KPZ fixed point.

Let us first comment on the inviscid scale.
In both the $p_0$-DE and LO approximation, the length $L_{\nu=0}(\hat D_\Lambda)$ saturates to a plateau as $\hat D_\Lambda\to 0$. In the $p_0$-DE case, the values of the plateau are indistinguishable for different regulator parameters, and $\nu_\kappa$ crosses zero almost immediately after the beginning of the FRG flow, that is, at $L$ close to $\Lambda^{-1}$. This shows that the IB regime sets in already at relatively small scales,    and the intermediate range of scales where the IB regime extends is essentially independent of the microscopic noise $D_\Lambda$ or equivalently of the microscopic effective coupling $\hat g_\Lambda$.  This justifies why the IB regime is indeed very robust, independent of the large-scale behavior (EW or KPZ) and of the stochastic or deterministic nature of the equation, as was observed in Ref. \cite{Vercesi2024}.

We now turn to the KPZ scale.
The $p_0$-DE result for $L_\text{KPZ}(\hat g_\Lambda)$ perfectly follows the law $L_\text{KPZ} = \text{const} / \hat g_\Lambda$ for $\hat g_\Lambda \lesssim 10^{-2}$, which is consistent with (\ref{eq:estimationsSneppen}) (the constant is non-universal and depends on the regulator). As mentioned previously, the LO  result is less reliable numerically for small $\hat g_\Lambda$, we show it for completeness. It  gives $L_\text{KPZ} \propto \hat g_\Lambda^{-a}$ with $a=0.66$ or $0.47$ if $L_\text{KPZ}$ is defined via $\hat g_\kappa$ or $\eta^D_\kappa$, which is in qualitative agreement (power-law increase), although not quantitative, with the $p_0$-DE result. Note that $L$ is measured in units of $\Lambda^{-1}$, which is defined as the microscopic scale. Therefore, one  cannot directly extract the precise absolute value of the scale at which one or another scaling regime begins (this would require to match the microscopic scales), but rather the ratio between these scales. Or, reciprocally, one can deduce the ratio of numerical noise levels in two different simulations, given the scale where the KPZ regime sets in: for instance, one estimates a $10^3$ ratio between the noise levels in Refs.~\cite{Hayot1993} and \cite{Roy2020}.

The most important outcome is that the behavior of $L_\text{KPZ}$ extrapolated to $D_\Lambda\to 0$ suggests that the KPZ regime only settles in the limit of infinite system size in the deterministic case. Therefore, it cannot be observed in principle in a finite system. Of course, in numerical simulations, computational noise is unavoidable, and it plays the role of a small $D_\Lambda$, rendering the KPZ regime observable, although very large sizes are required. This result is obtained within two reasonable approximations of the FRG, it would be interesting to find a rigorous proof of it.

\section{Conclusion}
\label{Conclusion}

We revisited the RG approach to the stochastic KS equation  using the functional renormalization group.
We first demonstrated that, within the Wilsonian RG, the flow equations for the KS model are not well-defined in the limit of sharp cutoff, which invalidates perturbative approaches based on it. To deal with the KS equation requires a smooth regulator,  as we implement in the FRG framework. The problem of using the sharp cutoff when momemtum-derivatives are needed is well-known is the field-theoretical literature, as well as the solution which lies in the use of a smooth cutoff. 
We derived the well-defined flow equations for the KS equation using FRG, first in a simple DE approximation. We showed that the effective microscopic viscosity  flows from its microscopic negative value $\nu$ to a positive value, and time-reversal symmetry emerges during the flow, leading in the IR to the KPZ fixed point. This thus establishes on firm  theoretical grounds the belonging of the KS equation to the KPZ universality class.

We then explained that, the usual DE approximation, which consists in an expansion around zero momentum, is not the most appropriate scheme to study 
the KS equations since this dynamics primarily involves and mixes a whole  range of momenta including  unstable modes. Indeed, the simple DE does not allow us to probe smaller value than 
$\nu/D\lesssim-0.25$. Thus, we resort to a full functional description, accounting for  momentum and frequency dependent renormalized  functions, which is the NLO approximation. Solving the NLO flow equations within the \textit{two grids} numerical scheme, we obtain the two-point correlation function in a wide range of momenta and frequencies. We identify and characterize the three universal regimes of the KS equation: the superdiffusive KPZ regime at small momenta, the diffusive EW regime at large momenta and small frequencies, and the ballistic IB regime at large momenta and larger frequencies.

Finally, we study the deterministic limit letting the microscopic noise $D_\Lambda$ tend to zero to explore the onset of the KPZ regime. In contrast to the KS equation itself, where increasing the noise amplifies the intrinsic instability, a large noise regulates, in some sense, the instability of the FRG equations caused by the negative viscosity, and the NLO flow becomes numerically unstable when the noise level is too small. We proposed another approximation, the $p_0$-DE, which turns this problem into an advantage: an expansion around the instability point allows us to obtained  analytical asymptotic flow equations, which we integrated. We determined the length  $L_\text{KPZ}$ where the KPZ regime sets in as a function $D_\Lambda$, and we obtained   a power-law dependence as $L_\text{KPZ} = \text{const} / \hat g_\Lambda$. This result is consistent with  previous estimations from numerics. This suggests that the KPZ regime is never reached in a finite-size deterministic system, while the presence of noise  reduces the system size required to observe it. Of course, numerical simulations always involve some computational noise and, in any real physical system, some thermal noise (and possibly  also some non-thermal noise) is present, which renders the KPZ regime observable in these cases.

An open issue is whether the  stochastic term added to the KS equation  accurately models the chaotic nature of the KS dynamics itself, in other words, whether the statistical average over this noise is a faithful proxy for the average over random initial conditions. Moreover, investigating the nature of the trajectories in the limit of zero noise is highly non-trivial as it can lead to spontaneous stochasticity. Within the present study, we focused on average quantities such as correlation functions and effective parameters, but did not consider the probability distributions of trajectories. The emergence of spontaneous stochasticity
 have started to be investigated using RG in simple models \cite{Eyink2020, Mailybaev2025}, it would be interesting to extend these approaches to the KS equation.

\section{Acknowledgments}

NW thanks the LPMMC,  LG and LC the Instituto de F\'isica de la Facultad de Ingenier\'ia, for hospitality during the completion of this work, and they acknowledge support from the French-Uruguayan Institute of Physics (IFU$\Phi$). LG acknowledges support  by the MSCA Cofund QuanG (Grant Number : 101081458) funded by the European Union. Views and opinions expressed are however those of the authors only and do not necessarily reflect those of the European Union or Université Grenoble Alpes. Neither the European Union nor the granting authority can be held responsible for them.

\appendix

\section{Simplification of Eqs.~\eqref{eq:dsDd} and~\eqref{eq:dstau}}
\label{app:integration-by-parts}

The flow equations for $\zeta_{\kappa}$~\eqref{eq:dsDd} and $\tau_\kappa$~\eqref{eq:dstau} can be expressed in terms of $r$ and $r'$ only, eliminating $r''$, which is convenient since it then allows one to use the modified Litim regulator~\eqref{eq:regLitim}. 
For this, one has to properly take into account the boundary terms, which exactly cancel out the singular parts of the integrals. 
In details, one can rewrite combinations of the type $(r')^n r''$ as
$\left( (r')^{n+1} \right)'/(n+1)$.
Then, after the change of variable $x\equiv q^2$, the integrands consist of terms of the form
\begin{align}
t_{n,m,k,p}&\equiv \int_0^\infty dx \,x^n \,r(x)^m  \frac{\left((r'(x))^k\right)'}{h_\nu(x)^p} \nonumber\\
& = \tilde t_{n,m,k,p} - \int_0^\infty dx \, (r'(x))^k\, \left(\frac{x^n r(x)^m}{ h_\nu(x)^p} \right)'
\end{align}
where $\tilde t_{n,m,k,p}$ denotes the boundary term.
It is zero if $2k+m<n+p$, while it is singular at the lower boundary $x=0$ if $2k+m\geq n+p$, and given by
\begin{equation}
\tilde t_{n,m,k,p} = -(-1)^k(\beta\gamma)^{k+m}(\alpha\beta\gamma)^{-6}\left( x^{n+p-m-2k} \right)'\Big|_{x\rightarrow 0} \,.
\end{equation}
In the integrands of (\ref{eq:dsDd}), (\ref{eq:dstau}) the only non-zero boundary terms correspond to $\tilde t_{1/2,3,2,6}$ and $\tilde t_{-1/2,4,1,6}$, both of them being of  order $x^{-3/2}$. They cancel out exactly with $x^{-1/2}$ terms entering the integrands which when integrated yield  $x^{-3/2}$ contributions compensating exactly the singular boundary terms. The resulting equations are available in \cite{supmat_nb} as a Mathematica notebook.

\section{Discussion of the {\it flow-A} adimensionalization}
\label{app:flowA}

In this appendix, we further discuss the {\it flow-A} choice of adimensionalization, and argue in particular that it does not ``enforce'' the fixed point to have KPZ scaling.
Indeed, another value of $\eta^A$ can be chosen, provided that the regulator (\ref{eq:regFlowA}) indeed  regulates the pole (it turns out that the regulator is strong enough when $\eta^A\gtrsim1/2$ for our choice of the regulator shape). The choice $\eta^A\neq1/2$ leads to a constant drift of the dimensionless functions $\hat f^{D,\nu}_\kappa$ and coupling  $\hat g^A_\kappa$ at the fixed point instead of a constant shape of $\hat f^{D,\nu}_\kappa$ or a plateau for $\hat g_\kappa$, while the exponents reach the plateau value 0.5 whatever the value of $\eta^A\gtrsim1/2$).

Another way to verify that the KPZ fixed point is reached in an unbiased manner is to replace the regulators (\ref{eq:regFlowA}) with the ``usual'' ones
\begin{align}
R_\kappa^D(p^2) = D_\kappa r_D(\hat p^2),\,
R_\kappa^\nu(p^2) = \nu_\kappa r_\nu(\hat p^2),\,
r_\nu=r_D \label{eq:regNLOusual}
\end{align}
as soon as $\nu_\kappa$ is positive. Let us denote $\kappa_0$ such a scale. At this scale, one has $\nu_{\kappa_0}=C_\nu A_{\kappa_0}$. Thus, one can continue the \textit{flow-A} with the usual flow, initialized with the functions obtained from \textit{flow-A}  at $\kappa=\kappa_0$: $f^{D,\nu}_{\text{IC}}(\omega,p)=f^{D,\nu}_{\kappa_0}(\omega,p)$.
For the dimensionless functions this amounts to 
$\check f^D_{\text{IC}}(\check\omega,\hat p)=\hat f^D_{\kappa_0}(\hat \omega,\hat p)$
and
$\check f^\nu_{\text{IC}}(\check\omega,\hat p)=\hat f^\nu_{\kappa_0}(\hat \omega,\hat p)/C_\nu$, where the caron denotes adimensionalization by $\nu_\kappa$ instead of $A_\kappa$. This ensures that the regulator and the dimensionful functions are continuous at $\kappa_0$ and that the fixed point is achieved by the flow parameters independently. We checked that it is indeed the case. This continuation is used to obtain the data shown in Fig.~\ref{fig:L_of_g}.

\section{$p_0$-DE approximation}
\label{app:p0-LPA}

Within the NLO approximation, when the  initial noise $D_\Lambda$ is too small (or equivalently $\hat g_\Lambda$ is too small), the flow has an instability caused by the negative part of $\hat f^\nu_\kappa(\hat \omega, \hat p)$ for the following reasons. In fact, this negative area in the $(\hat \omega, \hat p)$-plane moves from small-$\hat p$ region at $\kappa \sim \Lambda$ to larger $\hat p$ at intermediate RG times when $\nu_\kappa$ crosses zero. Even when $\nu_\kappa$ is already positive (that is, $\hat f^\nu_\kappa(\hat \omega=0, \hat p=0)>0$), the function $\hat f^\nu_\kappa(\hat \omega, \hat p)$ still has a negative part at large $\hat p$ at any $\hat \omega$, as can be observed  in  Fig.~\ref{fig:flow}(b). 

\begin{figure}
\begin{overpic}[percent,width=0.3\textwidth]{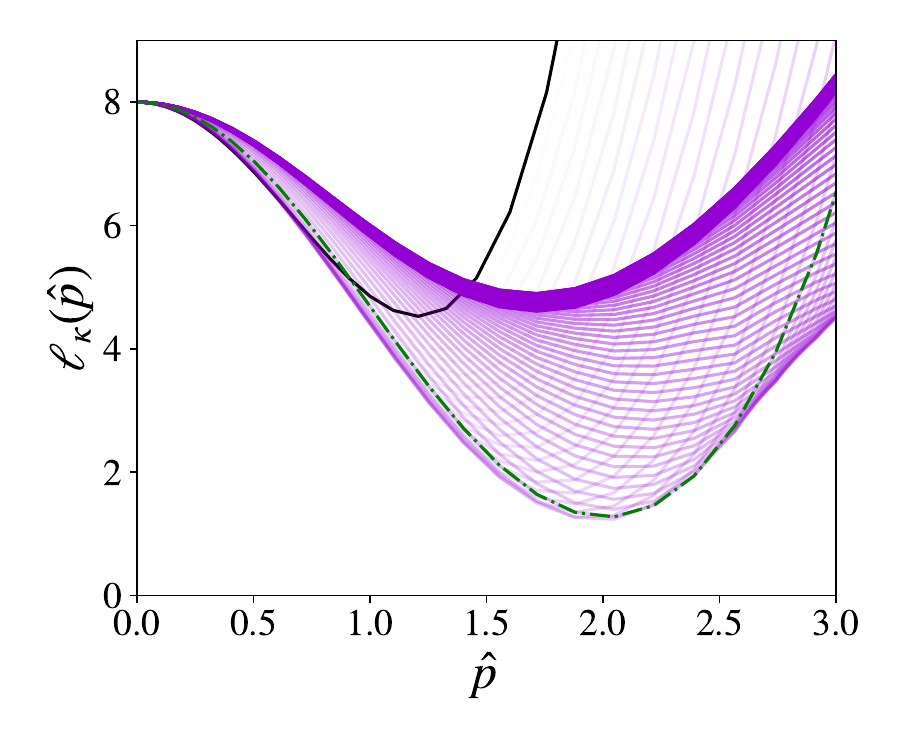}
    \put(82,22){(a)}
\end{overpic}
\begin{overpic}[percent,width=0.45\textwidth]{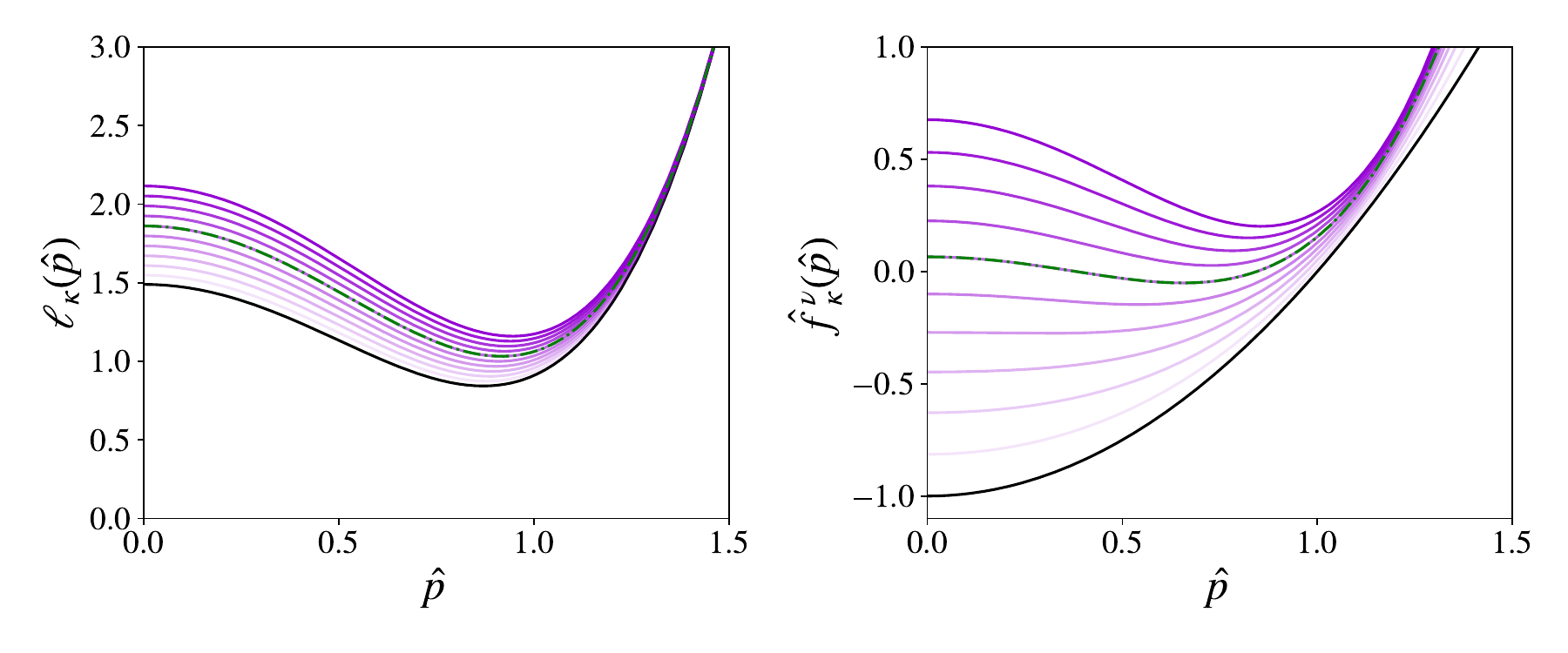}
    \put(39,12){(b)}
    \put(90,12){(c)}
\end{overpic}
  \caption{(a): Beginning of the flow of the denominator of the regulated propagator  $\hat{\ell}_\kappa$ within the {\it flow-A} NLO approximation; (b) and (c): beginning of the flow of $\hat{\ell}_\kappa$ and $\hat{f}^{\nu}_\kappa$ within the $p_0$-DE approximation. The black line marks the initial condition, darker violet lines indicate later RG times (larger length scales), the green dash-dotted line highlights the scale at which the effective viscosity just crosses zero.
  }
  \label{fig:flow_l}
\end{figure}

As a consequence, the denominator of the regulated propagator has a minimum, and this plays a very important role in the KS dynamics. This include
denominator at $\omega=0$ reads 
$\ell_\kappa(\hat p) \equiv  [\Gamma^{(1,1)}_{\kappa}(\omega=0,p)+R^\nu_\kappa(p)]/(\kappa^2A_\kappa) = \hat p^2(\hat f_\nu(0, \hat p)+R^\nu_\kappa(p)/A_\kappa)$
and is displayed in Fig.~\ref{fig:flow_l}(a). It shows a deep minimum
at a certain $\kappa$-dependent point, that we denote $\hat p_{0\kappa} \equiv \text{argmin}_{\hat p} \ell_\kappa(\hat p)$. The function $\ell_\kappa(\hat p)$ appears in the denominator of the integrands of $\hat I^{D,\nu}_\kappa(\hat\omega,\hat p)$ in (\ref{eq:flowAdimlessEqs}). In the example of the figure, the noise level is large enough for the flow to be stable ($\hat g_\Lambda=1$), but at smaller $\hat g_\Lambda$ the minimum approaches zero, and a very small denominator produces numerical instability.

In fact, this observation inspires an alternative approximation. Indeed, the presence of this minimum can be exploited to conceive an accurate approximation to capture the dynamics at small noise. The idea is to perform a DE expansion, but around the non-trivial $\hat p_{0\kappa}$ instead of zero momentum. It turns out that  $\hat I^{D,\nu}_\kappa(\hat\omega,\hat p)$ can be integrated  analytically if one assumes $\text{min}_{\hat p} \ell_\kappa$ to be very small.  In fact, a similar phenomenon occurs when the low-temperature behavior of magnetic systems is analyzed as a function of an external field (rather than as a function of momentum as in the present problem) in the context of the DE. A similar strategy has been explored to address the analysis of such systems \cite{Berges2002,Pelaez:2015nsa}. We call this approximation $p_0${-DE}. It is all the more accurate that the  $\text{min}_{\hat p} \ell_\kappa$  is small, which is precisely what occurs in the limit of small noise. Thus, it is very complementary to the NLO approach, as it allows one to investigate the small noise limit where the NLO approximation becomes less stable.

In this section, we neglect the $\omega$-dependence of the flowing functions (LO approximation), which is sufficient for the purpose of Fig.~\ref{fig:L_of_g}. The integrals $\hat I^{D,\nu}_\kappa(\hat\omega=0,\hat p)$, after performing  analytically the frequency integrals, consist of terms of the form 
\begin{align}
I_{klmn}(\hat p) = \int_{-\infty}^{+\infty} \frac{d\hat q}{2\pi}
\frac{ \hat q^2\,s_{klmn}(\hat p,\hat q)\ell_\kappa^k(\hat p+\hat q)}
{\ell_\kappa^l(\hat q)\ell^m_\kappa(\hat p+\hat q)\big[\ell_\kappa(\hat q)+\ell_\kappa(\hat p+\hat q)\big]^n}
\label{eq:s1}
\end{align}
($k\in\{0,1\}$, $l,m,n\in\{0,1,2\}$). Note that $\ell_\kappa(\cdot) = \ell_\kappa(|\cdot|)$.
At $\hat p = 2\hat p_{0\kappa}$, a ``resonance'' of the two minima occurs when the integration variable reaches a value $\hat q = -\hat p_{0\kappa}$. This affects the integrals $\hat I^{D,\nu}_\kappa(\hat\omega,\hat p)$ in the rhs of the flow equations (\ref{eq:flowAdimlessEqs})  and results in a hump of $\hat f^D_\kappa$, visible in Fig.~\ref{fig:flow}(a) at $\hat p \simeq 2 \hat p_{0\kappa}$ ({\it eg} for the green line, $\hat p_{0\kappa}\simeq 2$ and the hump is seen at $\hat p\simeq 4$).
Thus we expand the function $\ell_\kappa$ around $\hat p_{0\kappa}$ as
\begin{equation}
\ell_\kappa(\hat q) =\ell_{0\kappa}+\ell_{2\kappa}(\hat q^2-\hat p_{0\kappa}^2)^2 \,.
\label{eq:l_approx}
\end{equation}
This corresponds to
\begin{equation}
\hat f^\nu_\kappa(\hat q)= a_\kappa+b_\kappa(\hat q^2-\hat p_{0\kappa}^2)+c_\kappa(\hat q^2-\hat p_{0\kappa}^2)^2\,,
\label{eq:l-exp}
\end{equation}
where the coefficients $a_\kappa,b_\kappa,c_\kappa$  are easily recovered from $\ell_{0\kappa}$, $\ell_{2\kappa}$, and $\hat p_{0\kappa}$. The function $f^D_\kappa(\hat p)$
 is simply approximated by $f^D_\kappa(\hat p)=D_\kappa$.
 As a consequence, the integrand
\begin{align}
\int_{-\infty}^{+\infty} \frac{d\hat q}{2\pi} \frac{s(\hat p,\hat q)}{\ell_\kappa(\hat q)} \simeq
\int_{-\infty}^{+\infty} \frac{d\hat q}{2\pi} \frac{s(\hat p,\hat q)}{\ell_{0\kappa}+\ell_{2\kappa}(\hat q^2-\hat p_{0\kappa}^2)^2}
\end{align}
 behaves as a Lorentzian, all the more peaked that $\varepsilon_\kappa=\ell_{0\kappa}/\ell_{2\kappa}$ is small.
Under this assumption, we can calculate the integrals using analytical Lorenzian integration. For a small $\varepsilon$, and a smooth function $f(x)$, the following approximation holds:
\begin{align}
\int_{-\infty}^{+\infty} dx \frac{\varepsilon^2 f(x)}{\varepsilon^2+(x-x_0)^2}&=
\int_{-\infty}^{+\infty} dx \frac{f(x)}{1+(\frac{x-x_0}{\varepsilon})^2} 
\nonumber\\
&=\varepsilon\int_{-\infty}^{+\infty} du \frac{f(\varepsilon u+x_0)}{1+u^2} \nonumber\\
&\simeq
\varepsilon\int_{-\infty}^{+\infty} du \frac{f(x_0)}{1+u^2} = \varepsilon f(x_0) \pi\,
\end{align}
where  $f(x)$ is  such that the integral is convergent.

In practice, the integrands in (\ref{eq:s1}) have a more complicated form with a product of $\ell_\kappa$'s evaluated at different momentum configurations. In fact,  $1/\ell_\kappa(\hat q)$ has 2 peaks at $\hat q=\pm \hat p_{0\kappa}$, and $1/(\ell(|\hat q|)+\ell(|\hat p+\hat q|))$ has only one peak at $\hat q=-\hat p_{0\kappa}$ (other roots are imaginary). The whole expression thus has two peaks $\hat q=\pm\hat p_{0\kappa}$, but the dominant one is $\hat q=-\hat p_{0\kappa}$.
 We hence approximate integrals such as \eqref{eq:s1} as
\begin{align*}
I_{klmn}(2\hat p_{0\kappa}) &\simeq a_{klmn} \int_{-\infty}^{+\infty} \frac{d\hat u}{2\pi}
\frac{\hat p_{0\kappa}\,s_{klmn}(2\hat p_{0\kappa}, -\hat p_{0\kappa})}{2\big(1+\frac{(\hat u-\hat p_{0\kappa}^2)^2}{\varepsilon_{\kappa}}\big)^{l+m+n-k}}
\,,\\
a_{klmn}&=\dfrac{1}{2^n\ell_0^{l+m+n-k}}
\end{align*}
and analytically perform the integral over $\hat u$ (which is given in terms of simple $\Gamma$ functions).
Moreover, some flows involve the derivative with respect to $\hat q^2$ of $\ell_\kappa(\hat p+\hat q)$ (see \eqref{eq:dsl0l2p0} below). In this case, this derivative is  replaced with the derivative of $\ell_\kappa(\hat q)$, which is given by $2\ell_{2\kappa}(\hat q^2 -p_{0\kappa}^2)$ in the expansion \eqref{eq:l_approx}. The remaining $\hat q$-dependence is neglected, {\it i.e.} $\hat q$ is replaced by $-\hat p_{0\kappa}$.

This approximation correctly captures the behavior of the integrals when $\varepsilon_\kappa$ is small. It allows us to evaluate $\hat I^{D,\nu}_\kappa(\hat\omega=0,\hat p)$ at $\hat p=2\hat p_{0\kappa}$ analytically with high accuracy
(benchmarks vs numerical integration at given parameters values can be found in \cite{supmat_nb}).

Thus, the flow equations of $\Gamma_\kappa^{(0,2)}$,  $\Gamma_\kappa^{(1,1)}$  and $\p_{\hat p^2}^n\Gamma^{(1,1)}_\kappa$ can be efficiently evaluated at $\hat\omega=0$ and $\hat p=2\hat p_{0\kappa}$.  The analytical result is a function of $\hat p_{0\kappa}$, $\ell_{0\kappa}$, $\ell_{2\kappa}$,  and one can derive from them the flows  of $\ell_\kappa$, $\p_{\hat p^2}^n \ell_\kappa$ and $\eta^D_\kappa=-\p_s \ln D_\kappa$ also evaluated at $\hat p=2\hat p_{0\kappa}$. Assuming that the expansion (\ref{eq:l-exp}) around $\hat p=\hat p_{0\kappa}$ is still accurate at $\hat p=2\hat p_{0\kappa}$, we can find the flow equations of the expansion parameters as:
\begin{align}
&\p_s \ell_{0\kappa} = \big[
    \p_{s}\ell_\kappa(\hat p) + 
    \frac{9 \hat p_{0\kappa}^4}{2} \p_{\hat p^2}^2 \p_{s}\ell_\kappa(\hat p) - 
    3 \hat p_{0\kappa}^2\p_{\hat p^2}\p_{s}\ell_\kappa(\hat p)
\big]\Big|_{\hat p = 2\hat p_{0\kappa}}\nonumber\\
&\p_s \ell_{2\kappa} = \frac{1}{2} \, \p_{\hat p^2}^2 \p_s \ell_\kappa(\hat p) \Big|_{\hat p = 2\hat p_{0\kappa}} \nonumber\\
&\p_s \hat p_{0\kappa}^2 = \frac{3 p_{0\kappa}^2 \p_{\hat p^2}^2 \p_s \ell_\kappa(\hat p) - \p_{\hat p^2} \p_s \ell_\kappa(\hat p)}{2\ell_{2\kappa}}\Big|_{\hat p = 2\hat p_{0\kappa}}\,.
\label{eq:dsl0l2p0}
\end{align}
The explicit form of these flow equations and their derivation can be found in \cite{supmat_nb}. The result of their numerical integration is displayed in Fig.~\ref{fig:flow_l}(b,c).

The $p_0${-DE} approximation is valid when $\ell_{0\kappa}/\ell_{2\kappa}$ remains small, that is, in the beginning of the RG flow, when $\nu_\kappa$ is negative or small positive, say, $\nu_\kappa<c_\nu$. At $\kappa=\kappa_0$ such that $\nu_{\kappa_0}=C_\nu<c_\nu$ we switch the flow equations to the usual simple DE (expanded around zero-momentum), suitable for the KPZ equation with parameters $\nu_\kappa>0$  and $D_\kappa$.
Due to the time-reversal symmetry of the KPZ equation, the FDT relation (\ref{eq:FDT1}) leads to $\nu_\kappa =  \alpha D_\kappa$ and $\eta^D_\kappa=\eta^\nu_\kappa\equiv\eta_\kappa$. 
Thus, at $\kappa<\kappa_0$,  the flow is described by only two parameters: $\hat g_\kappa = \kappa^{-1} \lambda^2 D_\kappa / \nu_\kappa^3$ and $\eta_\kappa$. As in 
 Appendix~\ref{app:flowA}, to ensure continuity of the regulator and the (dimensionful)  flow parameters, we choose in $p_0$-DE a regulator in the form (\ref{eq:regFlowA}) ($\alpha = C_\nu$ in this case). The integration of the $p_0$-DE flow and its continuation with the usual DE flow is also available in \cite{supmat_nb}.


%
\end{document}